\documentclass[twocolumn, switch]{article} 
\pdfoutput=1
\usepackage{preprint}

\usepackage{booktabs} 
\usepackage{graphicx}		
\usepackage{subcaption}
\usepackage{tabularx}
\usepackage{amsmath,amsfonts,amssymb,amsthm}

\usepackage[backend=biber,natbib=true,maxbibnames=99]{biblatex}

\usepackage[utf8]{inputenc} 
\usepackage[T1]{fontenc}  
\usepackage{xcolor}   
\usepackage{booktabs}     
\usepackage{microtype}    
\usepackage{lineno}   
\usepackage{float} 

\usepackage{titlesec}
\titlespacing\section{0pt}{12pt plus 3pt minus 3pt}{1pt plus 1pt minus 1pt}
\titlespacing\subsection{0pt}{10pt plus 3pt minus 3pt}{1pt plus 1pt minus 1pt}
\titlespacing\subsubsection{0pt}{8pt plus 3pt minus 3pt}{1pt plus 1pt minus 1pt}

\title{Making root cause analysis feasible for large code bases: a solution approach for a climate model}

\author{
  Daniel J.~Milroy \\
  University of Colorado Boulder \\
  \texttt{daniel.milroy@colorado.edu}\\
   \And
 Allison H.~Baker \\
  \small{National Center for Atmospheric Research}\\
  \texttt{abaker@ucar.edu} \\
  \And
 Dorit M.~Hammerling \\
  \small{National Center for Atmospheric Research}\\
  \texttt{dorith@ucar.edu} \\
  \And
 Youngsung~Kim \\
  \small{National Center for Atmospheric Research}\\
  \texttt{youngsun@ucar.edu} \\
  \And
 Elizabeth R.~Jessup \\
  University of Colorado Boulder \\
  \texttt{elizabeth.jessup@colorado.edu} \\
  \And 
 Thomas~Hauser \\
  University of Colorado Boulder \\
  \texttt{thomas.hauser@colorado.edu} \\
}

\begin{document}

\twocolumn[ 
  \begin{@twocolumnfalse} 
  
\maketitle

\begin{abstract}
Large-scale simulation codes that model complicated science and 
engineering applications typically
have huge and complex code bases. 
For such simulation codes, where bit-for-bit comparisons 
are too restrictive, finding the source of statistically 
significant discrepancies (e.g., from a previous version, 
alternative hardware or supporting software stack) in 
output is non-trivial at best. Although there are 
many tools for program comprehension through debugging 
or slicing, few (if any) scale 
to a model as large as the Community Earth System Model 
(CESM\texttrademark), which consists of more than
1.5 million lines of Fortran code.
Currently for the CESM, we can easily determine 
whether a discrepancy exists in the output using a by now well-established 
statistical consistency testing tool.  However, 
this tool provides no information as 
to the possible cause of the detected discrepancy, 
leaving developers in a seemingly impossible 
(and frustrating) situation. Therefore, our aim 
in this work is to provide the tools to enable 
developers to trace a problem detected through the 
CESM output to its source. To this end, our strategy 
is to reduce the search space for the root 
cause(s) to a tractable size via a series of 
techniques that include creating a directed 
graph of internal CESM variables, extracting a subgraph 
(using a form of hybrid program slicing), partitioning 
into communities, and ranking nodes by centrality.  
Runtime variable sampling then becomes feasible 
in this reduced search space. We demonstrate the utility 
of this process on multiple examples of CESM simulation 
output by illustrating how sampling can be performed as 
part of an efficient parallel iterative refinement procedure to 
locate error sources, including sensitivity to CPU instructions. 
By providing CESM developers with tools to identify 
and understand the reason for statistically distinct output, 
we have positively impacted the CESM software development 
cycle and, in particular, its focus on quality assurance.
\end{abstract}
\vspace{0.35cm}

  \end{@twocolumnfalse} 
] 



\section{Introduction}
\label{sec:intro}
The Community Earth System Model (CESM\texttrademark) 
is a commonly used application for simulating 
the climate system, and its influence extends from science to policy. 
The model's Fortran code base is modular, 
which facilitates its evolutionary and 
community development. 
The CESM has grown to approximately 
1.5 million lines of code, which contain expressions of modern 
coding techniques together with code written in its earliest 
versions (decades ago).  CESM's size, complexity, and 
continuous development make finding sources of error and 
value discrepancy difficult. While there are many debuggers  
capable of locating causes of runtime errors and 
segmentation faults in large-scale applications, there are 
few tools designed for root cause analysis of value
discrepancies generated in large models 
(particularly those written in Fortran). We focus on the 
CESM in this work, though our root cause analysis 
methods may be applicable to 
other large Fortran models, or with a different 
parser, models written in other languages.

The first step to finding sources of 
inconsistency is to identify abnormal 
output. A simple test like bit-for-bit (BFB) 
equivalence is not useful because legitimate 
changes or optimizations to the model can result in bitwise 
differences between outputs. The works \cite{baker2015,baker2016} 
establish statistical testing for consistency 
with an ensemble of ``accepted'' output from the 
Community Atmospheric Model (CAM) and Parallel 
Ocean Program (POP) component models of CESM. 
Ensemble methods are common in weather forecasting 
and climate studies; using them to test 
experimental outputs for consistency is broadly 
useful unlike simple BFB equivalence. 
The Ensemble Consistency Tests (ECTs) 
quantify the internal climate model variability 
present in an ensemble of the respective 
component models' outputs. The ECT then evaluates 
new, experimental outputs in the context 
of the ensemble to determine whether the new outputs 
are statistically consistent. The tests (together 
referred to as the CESM-ECT) provide an 
objective measure of difference without resorting 
to excessively strict metrics like BFB results. 
Moreover, the output of the tests does not require expert
knowledge of climate science and returns a 
user-friendly \textit{Pass} or \textit{Fail}. 
The CESM-ECT has proven its value both in end-user 
testing and in port verification, where it was 
used to test CESM outputs from Cheyenne 
(the successor supercomputer) against the 
accepted ensemble generated on the Yellowstone 
supercomputer at the National Center for Atmospheric Research (NCAR). 

While the CESM-ECT has been shown to work very 
well for correctly classifying new outputs, 
in the case of a failure it provides no 
information on the location or nature of the root causes. 
In fact, our work here is motivated by the need to provide this crucial information, as its 
utility became apparent during our recent investigation into statistically distinct
CESM output 
between two large supercomputers (detailed in \cite{milroy2016}). Given CESM's large and complicated Fortran code base, 
determining the reason for this CESM-ECT \textit{Fail} required equal measures of data analysis, climate science knowledge, 
experience with the code base, and intuition. 
In all, we spent several months discovering the cause, and the process took the combined expertise of many 
scientists and engineers.
Clearly, automating this process would be a tremendous asset 
for software engineers and scientists and would accelerate 
CESM development.  In this work we make significant progress toward 
automating root cause analysis for sources 
of error and discrepancy in CESM, which is too 
large a model for direct application of currently available  
techniques.



This work is organized as follows.  In Section \ref{sec:method-related}, 
we overview our strategy and contributions 
and discuss related work. Section \ref{sec:varselect} 
describes identifying output 
variables most affected by inconsistencies.
In Section \ref{sec:ast-to-digraph}, we detail 
transforming approximately 660,000 lines of 
code into a directed graph (digraph). 
In Section \ref{sec:graph-analysis-method}, 
we define our method of iterative convergence 
to locate sources of discrepancy, and in 
Section \ref{sec:experiments}, we present examples of our method.

\section{Overview and related work}
\label{sec:method-related}

In this section we provide a summary of 
the methods that we develop and 
describe our principal contributions. We then
discuss related work on program slicing 
and runtime sampling.

\subsection{Method and contributions}
\label{subsec:method}
Here we overview our method and techniques for 
reducing the search space of possible causes of discrepancy. 
Recall that the process begins when CEST-ECT issues a \textit{Fail}, indicating a statistical difference (or discrepency) between the experimental output of interest and the accepted ensemble.
We want to identify statistical differences as early as possible (i.e., at an early time step) in the simulation for several reasons: bugs or discrepancies may not propagate changes through the entire model, climate feedback mechanisms may not yet take effect, and less of the source code is executed. Therefore, we use the CESM-ECT tool that evaluates consistency at time step nine, which is called "ultra-fast" CAM-ECT, or UF-CAM-ECT \cite{milroy2018}. 

A straightforward first step in reducing the search space is to eliminate modules that are not built into the final executable, and there may be many as CESM can be compiled with numerous configurations.  Then,
we use an existing code coverage tool to discard modules 
that are not yet executed by the second time step, as well as to remove unexecuted subprograms from the remaining modules.
Next, we further reduce the scope by determining the CESM output variables (i.e., those written to file) that are most affected by the discrepancy, allowing  us to disregard locations that compute other variables. In the commonly used CESM configuration that we use for our experiments in this work, these initial steps reduce the potential lines to search from about 1.5 million to 660,000, which is still 
substantial.

\begin{figure*}[t]
\includegraphics[width=\textwidth]{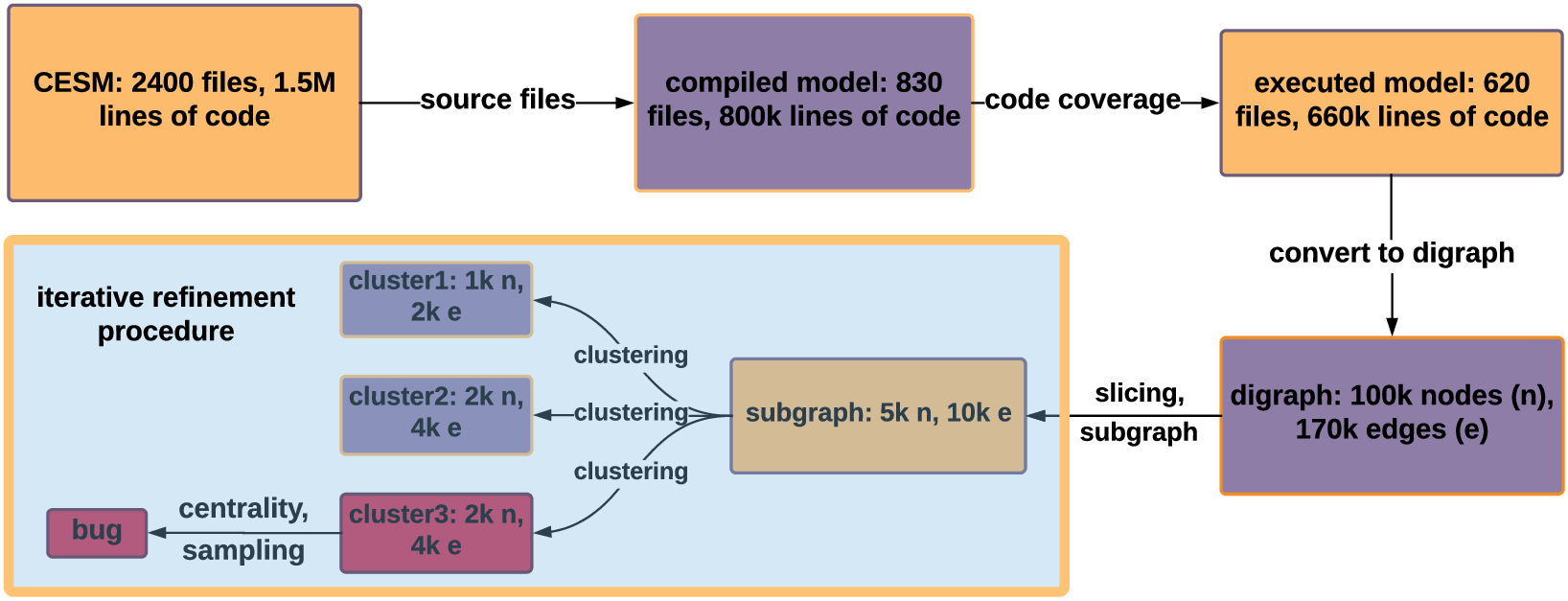}
\caption[paper process]
{Process flow schematic of our methods.}
\label{fig:process}
\end{figure*}

From this reduced code base, we construct a digraph of variable dependencies expressed through assignment statements. We then 
extract from this graph a subgraph that computes 
the variables that we previously identified as most affected by the discrepancy. Next, to facilitate parallelism and runtime sampling (among other benefits that we describe later), we use clustering to partition the subgraph. For each cluster, we rank nodes based on their centrality to determine which code variables to sample at runtime accordingly.
Finally, as a planned last step, we will further narrow the search space based on the runtime sampling results that indicate whether or not there are value differences between an ensemble and an experimental run, followed again by clustering and sampling by centrality to converge iteratively on the 
sources of discrepancy (currently performed in simulation). 
Figure \ref{fig:process} provides a schematic of our process.

While static program slicing, community detection, and centrality are well-studied and frequently used 
techniques in the areas of formal verification and graph analysis, to the best of our knowledge, their combination into a root cause analysis framework is novel. We make the following contributions in this work: 

\begin{itemize}
    \item We create a pipeline to convert the CESM source code into a digraph with extensive metadata that represents variable assignment paths.
    \item We develop a hybrid static program slicing approach that efficiently returns large slices.
    \item We devise an iterative refinement procedure based on community detection, centrality, and runtime sampling to contract the slice to a size amenable to analysis. Through community detection, the procedure can be performed in parallel.
    \item We perform experiments based on CESM output that demonstrate finding the causes of model discrepancy.
    \item We provide evidence that our methods accurately characterize information flow at runtime.
\end{itemize}

\subsection{Related work}
\label{subsec:relatedwork}

Program slicing is a common technique in debugging 
and in software development and maintenance that extracts 
sections of a program that can affect a particular 
region of code \cite{weiser1981,weiser1984}. In a broad sense, 
program slicing can be divided into two methods: 
static slicing, which considers all possible 
executions of a program, and dynamic slicing, 
which accounts for only one execution given 
a set of criteria (e.g., \cite{tip1994,silva2012}). Static slicing 
is generally less expensive but can return slices 
that contain too many extraneous statements 
to be useful \cite{bent2001}. Dynamic slicing 
can be far more precise but correspondingly expensive due 
to the inclusion of algorithms needed to evaluate 
the satisfiability of sections of the slice 
(such as SAT or Satisfiability Modulo 
Theory solvers \cite{harris2010}).  
So-called backward slicing considers subsets 
of code that affect a target location by 
backward traversal; it can be performed 
via static or dynamic slicing \cite{jaffar2012}. 
We are not aware of any dynamic slicing methods 
that can be applied to models consisting of over 
a million lines of code. We adopt the strategy of hybrid slicing 
\cite{gupta1995}, which uses dynamic information 
about program execution to refine static slices. 
In our case, the dynamic information is provided 
by a code coverage tool.

Program sampling or instrumentation provides 
detailed analysis of program states 
by monitoring variable values at 
runtime. This type of monitoring can be used to 
detect divergent values of individual variables 
but can be extremely expensive (both in space 
and time) depending on the sampling frequency 
and the number of variables monitored. Many 
debuggers and profiling toolkits can perform 
sampling of large, distributed-memory applications 
(Allinea MAP and DDT \cite{allinea2018}, TotalView 
\cite{totalview2018}, and Tau \cite{shende2006}, to name a few), and 
tools such as FLiT \cite{sawaya2017} 
and KGen \cite{kim2016,kim2017} can detect divergent values 
at runtime. We seek to reduce the search 
space of CESM to the point that such tools 
(or those of future design) can identify 
specific variables that cause model 
divergence.

\section{Identifying affected output variables}
\label{sec:varselect}

After UF-CAM-ECT returns a failure, our first step is to 
identify the CAM output variables that are 
affected (or most affected) by the cause of the failure. 
Doing so allows us to make a connection between 
the model outputs and the code itself.
Ideally, we perform a straightforward normalized comparison of floating 
point values at the first model time step, selecting only those variables 
that exhibit a difference between a single ensemble member 
and a single experimental run. Because this approach is the direct measure of difference, we 
recommend using it first.
However in our experience, it is most often the case that all CAM output variables are different 
at the model time step zero, so simply comparing floating point values is not
useful for narrowing down the number of variables. 
Therefore, for these cases, 
we instead examine properties of the variables' 
distributions with two variable selection 
methods to identify those most affected 
by the discrepancy.

The first method measures distances between 
the distribution medians of the ensemble and 
experimental runs for each variable. To make 
meaningful distance comparisons across variables, we 
standardize each variable's distribution 
by its ensemble mean and standard deviation. Then we identify 
variables whose interquartile ranges (IQRs) 
of ensemble and experimental distributions 
do not overlap. We then rank these variables 
by descending order of distance between their medians. Although this 
provides a straightforward ordering of variables, 
the disadvantage of this approach is that often many 
variables are identified. To be useful for tracing sources 
of inconsistency, the variable selection method 
should identify as few variables as possible 
(not more than 10). Our second method employs 
logistic regression with regularization via a penalized 
$L_1$-norm (known as the lasso). We generate a set of 
experimental runs and use this in conjunction with 
our ensemble set to identify the variables that best 
classify the members of each set. We tune the 
regularization parameter to select about five 
variables as that yields a subset of CESM and 
CAM that, in our experiments, contains the 
known source of statistical inconsistency while still being small.    
The variables selected by the lasso (and their order) 
mostly coincide with the order produced by 
computing the distance between standardized 
medians. 

CESM and CAM present 
a challenge for variable selection techniques, 
as the models' interconnectivity results 
in most changes (in software or hardware) 
propagating through the atmosphere model rapidly, 
and affecting most variables.  
Since many of the experiments in this work 
represent modifications to code present in the 
tightly connected ``core'' of CAM, it is reasonable 
that variable selection is difficult. 
Variable selection for smaller or 
simpler models may represent less of a challenge.

\section{From source code to digraph}
\label{sec:ast-to-digraph}
\begin{figure}[t]
\includegraphics[width=0.48\textwidth]{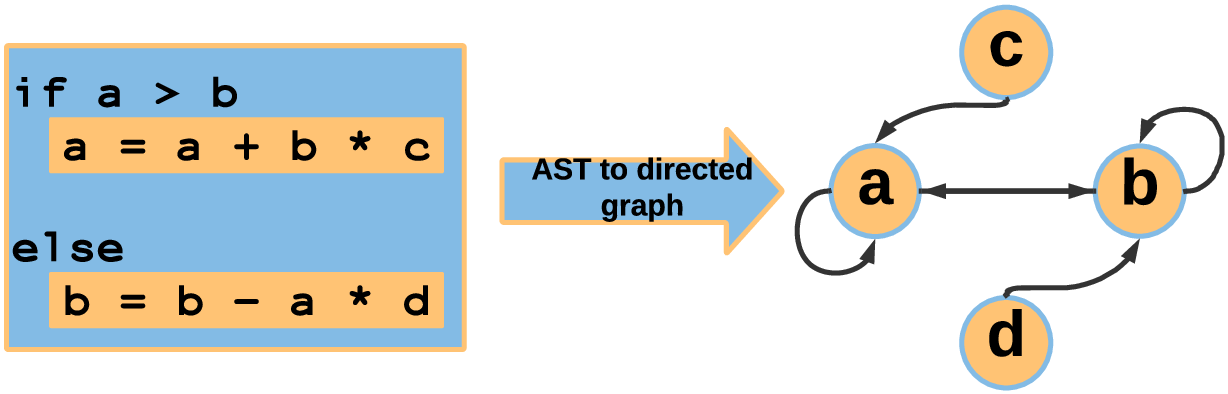}
\caption[Example code statement in three forms]
{Example statement as source code and its directed graph representation.}
\label{fig:src-dg}
\end{figure}

Finding lines of code that modify a particular CAM output variable 
seems a straightforward task: use a text-based 
search to select code that modifies the variable in question. 
However, many internal variables may alter values that eventually 
propagate to the affected output values, and the data 
dependencies are likely to be complicated. 
To describe the relationships between CESM variables accurately, 
we convert each source code file into an Abstract Syntax Tree (AST), 
which represents code syntax as structural elements 
of a tree. From the ASTs we create a digraph 
which represents variable dependencies.
Figure \ref{fig:src-dg} provides a simple example of 
the transformation of source code assignments to 
a digraph.

\subsection{Generating the AST}
\label{subsec:gen-ast}
To construct the AST for CESM, we need to 
parse the source code.  We 
use the same CESM version as in
\cite{kay2015}, and our experimental setup (FC5) consists of 
a subset of all available component models.
Before parsing, we do 
several preprocessing steps to exclude code that is 
not executed.
Unfortunately, the CESM build system 
obfuscates which components' Fortran 
modules are compiled into the specified model.
Therefore, we employ KGen \citep{kim2016,kim2017}, 
a tool to extract and run code kernels as standalone 
executables, to identify 
the files compiled into the 
executable model, reducing the number of 
modules from approximately 2400 
to the nearly 820 used by our experimental setup.
KGen also replaces preprocessor directives 
with their compile-time values, enabling 
conversion of Fortran code to 
a Python AST via fparser (based on F2PY \cite{peterson2009}). 
Fparser is the only tool that we are aware of to parse Fortran into 
Python data structures.

We further limit the scope of code considered 
by examining coverage, which identifies 
code lines, subprograms, and modules executed in a given 
application. Since our objective is to identify critical 
code sections as early as possible in the CESM runtime, 
we can ignore many subsections of code which are not yet 
run. To find such code, Intel provides a 
code coverage tool \cite{intelcodecov2017}
that writes profiling files that 
indicate coverage down to individual lines. In our 
experience, the tool returns accurate evaluations 
to the level of subprograms, but its behavior at the 
line-level is inconsistent.
Nevertheless, finding entire unused 
modules and uncalled subprograms is useful 
and reduces the number of modules and subprograms
to be parsed by about 30\% and 60\%, respectively. We develop 
software to parse the codecov HTML output, using the output to 
remove unnecessary modules 
and comment out unused subprograms.

\subsection{From AST to digraph}
\label{subsec:ast-digraph}

\begin{figure}[!t]
\includegraphics[width=0.475\textwidth]{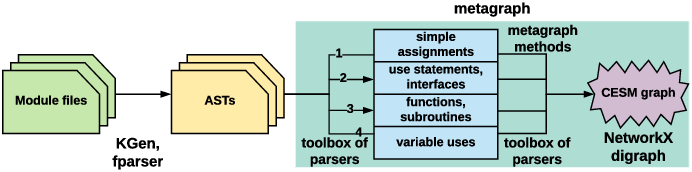}
\caption[paper process]
{Converting Fortran files into a metagraph.}
\label{fig:module-graph-flow}
\end{figure}

After converting each Fortran module file 
into an AST, we extract data dependencies 
to form a digraph.
See Figure \ref{fig:module-graph-flow} for a visual overview.
We need to resolve all assignments, as 
directed paths of assignments define dependencies 
between variables. Tracing dependencies between 
subprograms (similar to interprocedural 
program slicing \cite{weiser1984}) requires 
processing subroutine and function calls, 
interfaces, use statements, etc. 
Assignments without functions or arrays 
are processed immediately. 
To allow correct mappings between call 
and subprogram arguments, parsing 
statements with calls must be done after 
all source files are read. Furthermore, 
Fortran syntax does not always distinguish 
function calls from arrays, so correct 
associations must be made after creating a 
hash table of function names.

Transforming the CESM source code into a digraph presents 
several challenges. Fparser sometimes fails to convert 
a Fortran file into an AST due to bugs and
statements that exceed fparser's capabilities (e.g., 
one CESM file contains a statement with more than 3500 characters).
In fact, CESM contains thousands of expressions that are 
highly complex, with deep function and subroutine calls. Because existing 
Fortran parsing tools are inadequate for CESM, we employ three different 
parsers for each assignment (some are subjected 
to multiple passes of these parsers): fparser, 
KGen helper functions, and a
custom string parsing tool based on regular expressions 
and Python string manipulations that we developed 
to process cases unhandled by the other tools.

Processing the ASTs results in a 
\textit{metagraph} Python class that contains 
a digraph of internal variables, 
subprograms, and methods to analyze these structures. 
CESM internal variables are nodes with metadata, 
such as location (module, subprogram and line) and 
``canonical name'' (the variable name before 
being entered into the digraph - which requires 
unique node names). The digraph component 
of the metagraph is a NetworkX digraph \cite{schult2008}. 
NetworkX is a Python graph library that provides 
an extensive collection of 
easy to implement 
graph algorithms and analysis tools.

Note that with static analysis it is 
not always possible to determine which function 
a Fortran interface call actually executes at runtime.  
Therefore, we adopt the conservative approach of mapping all 
possible connections. We map the target of ``use'' statements to their local 
names to establish correct local symbols 
for remote procedures, resolving Fortran renames. 
If the use statement does not specify an ``only list,'' 
then we map all public variables in the source module to their 
target module variables. We do not consider chained 
use statements (i.e., where module A uses B, which 
uses C), since accurate dependency paths can be 
created by connecting the statements independently.  

With these associations 
defined, we iterate through statements 
containing subroutine calls and possible 
functions or arrays. We process 
subroutine and function calls by treating each argument 
as a tree, and we successively map outputs of lower 
levels to corresponding inputs above. 
Each output gets an edge to the above layer's input, which 
injects the call's graph structure into the CESM digraph. 
The top level argument output is connected to the subroutine's 
corresponding argument in its definition.  
Discerning functions from arrays is addressed by 
hash table lookups in the metagraph. 
Ultimately, the expression's right-hand-side variables and arrays 
and function (or subroutine argument) outputs are given edges to the left-hand-side. 
We process Fortran intrinsic procedures like \texttt{min} or \texttt{max}  
by creating paths from their inputs to themselves, and then to their outputs. 
We localize the intrinsic procedures to the line of code where they 
are called (i.e. a unique name like \texttt{min\_100\_\_modname}) to avoid 
creating spurious, highly connected variables. 
An example of node-edge mapping within a composite function 
is provided by the following process, where each 
function's internal variables form a path 
connecting its inputs to its outputs, in order of depth:

\begin{align*}
\omega &= \alpha(b(c, d)*e(f(g + h))) \nonumber \\
h &\rightarrow input(f) \nonumber \\
g &\rightarrow input(f) \nonumber \\
output(f) &\rightarrow input(e) \nonumber \\
c &\rightarrow input1(b) \nonumber \\
d &\rightarrow input2(b) \nonumber \\
output(e) &\rightarrow input(\alpha) \nonumber \\
output(b) &\rightarrow input(\alpha) \nonumber \\
output(\alpha) &\rightarrow \omega \nonumber \\
\label{eq:function-maps}
\end{align*}

We adopt a conservative approach for handling composite 
and complex Fortran data structures. Arrays are 
considered atomic in that we ignore indices. 
Pointers are treated as normal variables. 
Fortran derived types are challenging, 
as they can be chained into deep composite data 
structures. We define the indexed element of the 
derived type as the metagraph canonical name, 
e.g., \texttt{elem(ie)}
\texttt{\%derived}
\texttt{\%omega\_p} 
has a canonical name of ``omega\_p.''
In effect, we are compiling the CESM Fortran 
source code into node relationships in 
a digraph. Note that our parsing process is able 
to handle all but 10 assignment statements 
of the 660,000 lines of code in the coverage-filtered 
source. 

While details such as subroutine call handling, use 
statements, and derived types are specific to Fortran, 
the overall approach of converting source code to a directed 
graph can be accomplished with any compiled language.  In particular, 
the LLVM compiler infrastructure project \cite{lattner2004} features 
the Clang language front-end for C language family 
(C++, Objective C/C++, OpenCL, CUDA, and RenderScript) codes \cite{clang}.
When generating an intermediate representation of supported C language family code, 
Clang creates an AST with a convenient API (libclang). The libclang 
API provides Python bindings which permit AST traversal and 
source-AST mappings that can be used similarly to Fparser. 
We can adapt our Python software to traverse and convert 
the Clang AST to a directed graph for C/C++ codes as well. Modifying our 
project to enable analysis of C family codes is future work.

\section{Analyzing the CESM graph}
\label{sec:graph-analysis-method}

At this point, we have transformed the CESM code into a digraph that is composed of nodes, which 
are variables present in assignment expressions, 
and directed edges that indicate the directionality 
of the effect of one variable upon another. 
Now we narrow the scope of our search for the source of the discrepancy by 
analyzing the graph, usually accomplished by program slicing. 
Static slicing often produces slices 
that are too large to locate error sources, 
and dynamic slicing, while more precise, 
is too expensive to apply to the CESM graph
(about 100,000 nodes and 170,000 edges). 
Therefore, to make locating 
internal CESM variables or nodes 
that influence the values of the affected 
output variables more tractable, 
we examine static data dependency paths that  
terminate on these variables. 
We mitigate the imprecision of static backward slicing by 
integrating graph analysis algorithms to 
refine our slices. In this section, we 
discuss these methods and propose an 
iterative subgraph refinement procedure that involves runtime 
sampling of CESM nodes. \\

\subsection{Tracing affected internal variables}
\label{subsec:variables-in-graph}

Since variable relationships in assignment 
statements are represented as directed edges in the graph, 
we are interested in directed paths through CESM. 
These paths ignore control flow such as 
``if'' statements or ``do loops,'' 
so this approach is akin to backward 
static slicing. A key difference between 
our approach and typical program slicing is 
that nodes in the graph are single variables 
rather than expressions of multiple variables. 
Slicing criteria are thus single variables.
When used in conjunction with runtime information in the form of 
code coverage, our method can be considered a type of 
hybrid slicing (e.g., \cite{gupta1995}).

In NetworkX, the fastest way to 
determine dependencies is by computing shortest paths.
In particular, we seek the shortest paths that 
terminate on a CESM output variable. 
Unfortunately, finding the output variables in CESM is a challenge in 
its own right. Ideally, we would find the locations 
where I/O calls are made with the output variables 
as arguments, and then find all shortest paths in the graph
that end on those calls. Considering these paths does not work well in practice 
because CESM subprograms that write derived types, 
e.g., \texttt{state\%omega}, usually take the 
base derived type (\texttt{state}) as an argument, 
rather than the derived type element (\texttt{omega}). 
This means that there are few paths that terminate on 
\texttt{state\%omega} at the call location.
We address this problem by searching for paths 
that terminate on nodes with the canonical 
name (see Section \ref{sec:ast-to-digraph}) 
of \texttt{omega}.  This approach 
increases the size of our static slice, but with the 
attendant advantage that the discrepancy source will 
very likely be contained in the slice.

CESM I/O statements use
temporary variables extensively
and include character type variables in the 
output name argument, so 
uncovering the exact variable output for 
a given I/O call must be done with 
custom instrumentation. Of the nearly 1200 CAM I/O 
calls which write output variables, many 
include variables to label the output.
To resolve these variables, 
we instrument the code to print the corresponding 
string label, permitting a mapping 
between internal variable names and names 
written to file. For example, we do not search for 
paths that end on CAM output 
flds, but on variables whose canonical names 
are the internal name \texttt{flwds}.

So given a set of output variables that are 
affected by a certain change (determined as described in Section \ref{sec:varselect}), 
we compute the shortest 
directed paths that terminate on these variables 
with Breadth First Search (BFS). After finding these 
paths, we form the union of the node sets 
of all such paths. We are interested in the 
union rather than the intersection as multiple 
disjoint code sections can be involved in the 
computation of an affected variable. Such 
a scenario can arise when conditionals 
dictate whether I/O calls are 
executed. Using the union of all shortest 
paths terminating on the internal canonical names 
of affected output variables, we induce a 
subgraph on CESM, which yields the graph 
containing the causes of discrepancy.

\subsection{Community structure and node centrality}
\label{subsec:communities}

Since CESM and its component models are modular, 
it is reasonable to conclude that its graph 
should exhibit clusters corresponding to the 
modules or related processes. 
Induced subgraphs of CESM may contain 
cluster or community structure that can be 
exploited to improve our root cause search, 
which ends with sampling affected variables.
Since sampling can be an expensive process, 
only a limited number of nodes in the subgraph 
should be instrumented. By partitioning the 
subgraph through community detection, we can choose a small 
number of highly connected nodes in each community 
to sample and perform the instrumentation of these 
nodes independently (in parallel). This process 
can be performed iteratively to reduce the 
search space.

CAM contains two main processes: physics (sub-grid scale) 
and dynamics, which taken together feature a set of highly 
connected modules (the ``core''). These CAM modules are 
involved in the computation of many of the output variables, 
and the sources of the discrepancy (which we will also refer 
to as "bugs" for simplicity) are likely to affect multiple output variables.
An examination of node connectivity in the core 
reveals clustering of highly connected nodes 
in different communities. 
Although sampling the whole core's most connected nodes 
may detect floating point differences 
between ensemble and experimental runs, 
instrumenting highly connected nodes in each community instead
can reduce the distance between instrumented variables and bug 
locations (reducing the number of iterations needed 
to refine the search space).

Centrality is a fundamental way to distinguish 
nodes in a graph. Two simple examples of 
centrality are degree centrality, which counts 
the number of edges connected to a given node, and 
betweenness centrality, which counts the number of 
BFS or Dijkstra shortest paths (for weighted graphs) 
that traverse a node (or edge). 
Graph analysis via centralities proves useful in 
many diverse areas of research, e.g., 
\cite{freeman1978,salathe2010,shah2010,clauset2015}. 
A study of the relationship between brain regions' 
centralities and physical and cognitive function 
\cite{vandenheuvel2013} is particularly relevant to our work. 
They conclude that such analysis consistently 
identifies structural hubs (high centrality regions) 
in the cerebral cortex, and that 
``high centrality makes hubs susceptible to disconnection 
and dysfunction.''

The Girvan-Newman algorithm (G-N) \cite{girvan2002,newman2004} is 
a popular method for identifying communities in undirected 
graphs. The algorithm is based on edge betweenness centrality, 
which ranks edges by the number of shortest paths 
(computed via BFS) that traverse them.
The algorithm successively removes the edge with highest 
centrality in each connected component, which 
breaks the graph into ever smaller communities. 
G-N identifies communities via the following steps 
\cite{girvan2002}: 
\begin{enumerate}
    \item calculate the betweenness for all edges in the network 
    \item remove the edge with the highest betweenness
    \item recalculate betweenness for all edges affected by the removal 
    \item repeat from step 2 until no edges remain
\end{enumerate}
In practice each iteration involves removing the edge with 
the highest betweenness until the number of communities 
increases \cite{newman2004}. Note that G-N was 
formulated to identify communities in undirected graphs. 
In our case, we convert the directed 
subgraph into an undirected subgraph for purposes of 
community detection. This conversion is desirable for our work, 
as it is equivalent to forming the weakly connected 
graph of the directed subgraph. Weakly connected graphs 
are digraphs where any node can be reached 
from any other node by traversing edges in either 
direction. Bug locations may be anywhere in the subgraph, 
so we cannot impose assumptions about whether instrumented 
nodes are reachable via bug sources in the digraph 
(even between communities) in either direction. 
However, in our experiments (presented in Section \ref{sec:experiments}), we know where the bug 
locations are, so we can simulate 
how our sampling procedure detects 
floating point differences between the ensemble and experiment.  
Given our knowledge of directed paths' connectivity from 
known bug sources to central nodes, we can deduce whether a difference 
can be detected. For our method to be useful in situations 
where bug locations are unknown, we cannot assume such knowledge 
when we identify communities.

\subsection{Finding important nodes with centrality}
\label{subsec:centrality}
Given a modification that alters the values 
of a set of output variables, 
we seek locations in CESM that influence  
their computation. The CESM digraph lacks any 
information about the nature of connections 
between variables, so, for example, 
linear and exponential relationships are expressed 
identically in the graph. Indeed, 
the connectivity of the CESM graph is the only 
information we have to identify important locations 
in the code for sampling.

\begin{figure}[t]
\includegraphics[width=0.475\textwidth]{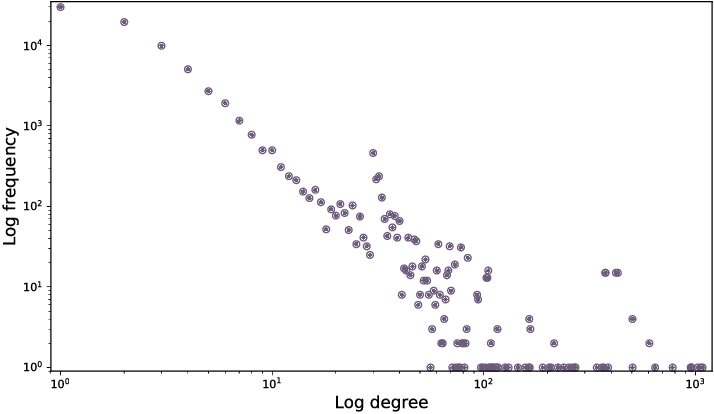}
\caption[CESM degree dist]
{Degree distribution of nodes in the CESM digraph.}
\label{fig:cesm-deg-dist}
\end{figure}

We use centrality to select nodes whose values 
are likely to be affected by the causes of 
statistical distinguishability. We can then sample 
the variables' runtime values to detect differences between 
an experimental and a control (or ensemble) run. 
Eigenvector centrality is a promising choice, as it 
considers not only the degree of each node, 
but the degrees of its neighbors and their neighbors, 
and is related to information flow in a graph. In fact, 
eigenvector centrality is related to 
PageRank, which is used to rank web pages 
in search results \citep{page1999}. 
In this work we focus on in-centrality, 
as we seek nodes which are likely 
to be affected by the bug sources. From 
the perspective of sampling, we are looking for information 
sinks rather than sources.
Eigenvector centrality has the disadvantage of favoring hubs (highly 
connected nodes), which ``causes most of 
the weight of the centrality to concentrate 
on a small number of nodes in the network'' 
for power law graphs \citep{martin2014}. 
The degree distribution of the total CESM graph 
approximately follows a power law, as can be seen in 
Figure \ref{fig:cesm-deg-dist}. Induced subgraphs 
of the CESM graph are also plausibly 
scale-free. A natural question is whether the concentration 
of centrality on graph hubs has undesirable 
effects on the ranking of nodes. 
We found that the application of non-backtracking centrality 
(based on the Hashimoto matrix \cite{hashimoto1989}) provides no advantage over 
standard eigenvector centrality for the CESM graph, 
its subgraphs, or communities. However, it may prove beneficial for 
models with graphs that follow a power 
law that produce more pronounced localization 
\citep{martin2014}.

\subsection{Iterative refinement procedure}
\label{subsec:iter-refinement}
Once communities are detected in the subgraph, we compute 
each community's eigenvector in-centralities and choose the 
top nodes to sample. The number of nodes to sample is
dictated by available computational resources. Based on whether 
a value difference can be found between the nodes sampled 
in the ensemble run and the experimental run, we can 
iteratively reduce the size of the subgraph to 
converge on the sources of statistical inconsistency. 
This iterative approach is similar to a $k$-ary search, which is a 
generalization of binary search. In binary search, 
the search space is halved and a single determination 
is made at each iteration, however for $k$-ary search 
the space is partitioned into $k$ sections and $k$ 
evaluations are made at each iteration. 
In our case $k$ varies by iteration depending on 
the number of communities identified. 
The following algorithm summarizes our overall approach:

\paragraph{Algorithm \ref{subsec:iter-refinement}}
\begin{enumerate} \label{alg:graphrefinement}
  \item Perform variable selection detailed in Section \ref{sec:varselect} \label{alg:i1}
  \item Map the set of affected CAM output variables in step 1 to their internal CAM 
    variables $\{V_i\}$ \label{alg:i2}
  \item For each affected internal variable $V_i$, use BFS to find the set of nodes $\{n_{i_j}\}$ in all 
    shortest paths that terminate on variables with canonical names 
    equal to $V_i$ in the CESM digraph \label{alg:i3}
  \item Form the induced subgraph $G$ via the union of nodes in the paths in step 3 \label{alg:i4}
  \item Use G-N to identify the communities $\{C_k\}$ of undirected $G$ (omitting 
    communities smaller than 3 nodes) \label{alg:i5} 
  \item Compute the eigenvector in-centrality for each $C_k$ and select $m$ nodes with largest centrality $\{n_{k_l}\}$ \label{alg:i6}
  \item Instrument $\{n_{k_l}\}$ for all $k$ in parallel for an ensemble run and 
    an experimental run, noting the set of nodes which take different values 
    $\{d_{k_l}\}$ ($\{d_{k_l}\} \subseteq \{n_{k_l}\}$) \label{alg:i7}
  \item 
    \begin{enumerate} 
      \item If $\{d_{k_l}\} = \emptyset$ (i.e., no different values are detected), form the \label{alg:i8a}
        induced subgraph on all nodes in $G$ that are not in BFS shortest paths that 
        terminate on $\{n_{k_l}\}$ 
      \item Else, form the induced subgraph of $G$ generated by nodes in $G$ that belong 
        to BFS shortest paths that terminate on $\{d_{k_l}\}$ \label{alg:i8b}
    \end{enumerate}
  \item Repeat steps 5-8 until the subgraph is small enough for manual analysis 
    or the bug locations are instrumented \label{alg:i9}
\end{enumerate}

There are three issues involved in the process above that 
merit discussion. First, it is possible that steps 
\ref{alg:i5}-\ref{alg:i8b} in algorithm \ref{alg:graphrefinement} 
do not refine the subgraph of 
the previous iteration (i.e., if the subgraph connectivity 
is such that all nodes are connected to all central nodes that 
take different values between the ensemble and experimental runs). 
In this case, we can select a subset of the most central nodes 
``most affected'' by the bugs. The second issue 
is that it is possible that the bug 
sources are not contained in any community i.e., if 
a bug is in an output variable that has only 
one neighbor. In this case, no different values will 
be detected in step \ref{alg:i7}, and the new induced subgraph 
will still meet the condition in step \ref{alg:i8a}. The next 
iterations will not detect differences, and the successive 
subgraphs will become increasingly disconnected. 
Eventually G-N will not identify 
any communities, and the resulting nodes will 
need to be analyzed. The third issue is an artifact of static 
slicing: since the paths do not take into account, e.g., 
conditional branches, some of the paths may not 
be traversed. We need to develop a method to track 
edge traversal and remove invalid paths; algorithm \ref{alg:graphrefinement}
must only remove nodes that actually can influence $\{n_{k_l}\}$
in step \ref{alg:i8a}.

Unless otherwise noted, we perform only one iteration of G-N 
in algorithm \ref{alg:graphrefinement} 
step \ref{alg:i5}. We could use a larger number 
to further subdivide the induced subgraph in each 
iteration (possibly enabling more parallelism), but 
we adopt a conservative approach to avoid 
clustering the subgraphs far beyond the 
natural structure present in the code. 
Note that excessive G-N iterations would not prevent algorithm 
\ref{alg:graphrefinement} from locating 
bug sources, but it can slow the process.

\section{Experiments}
\label{sec:experiments}

We apply the overall method discussed in 
Section \ref{subsec:iter-refinement}
to six experiments, and for the sake of brevity 
present four examples in detail. 
For all but one experiment, we introduce a bug 
into the source code so that the correct 
location is known. We then verify 
that our method can be used to identify 
the bug location in CESM by demonstrating 
how it would converge on the location 
given instrumentation. We first show that 
our method can correctly identify straightforward 
single-line bugs before proceeding to more 
complicated sources of output discrepancy,
such as the identification of variables most affected by 
certain CPU instructions. We make the 
following choices and assumptions in our 
experiments (unless otherwise noted): we restrict 
our subgraphs to nodes in CAM modules, 
perform a single G-N iteration (within algorithm \ref{alg:graphrefinement} 
step \ref{alg:i5}), 
choose the top 10 nodes by in-centrality to 
sample, and assume all paths are traversed at 
runtime. We restrict our experiments here to 
graph nodes from CAM modules as this reduces 
the number of iterations required and simplifies 
the graphs that we present. We note that we 
have successfully located bugs in the land module as well. 
Restricting nodes to CAM disconnects paths with segments in other component 
models, producing residual clusters of less than four nodes that  
are separated from the main communities. 
We remove these small clusters for the sake of clarity 
in our plots; their removal does not affect the results. 

In the figures that follow, subfigures \textbf{a} 
are the outputs of algorithm \ref{alg:graphrefinement} 
step \ref{alg:i4}, \ref{alg:i8a}, or \ref{alg:i8b}, 
depending on the iteration or whether simulated sampling detects differences.
Subfigures \textbf{b} color members 
of each community detected by step \ref{alg:i5}, 
and subfigures \textbf{c} represent the output of 
step \ref{alg:i7} for the community containing 
the discrepancy sources. 
Each of the following subsections describes a different experiment. 
Since we are unaware of methods capable of analyzing 
models as large as CAM or CESM, we do not perform comparisons 
with existing techniques.

\subsection{WSUBBUG}
\label{subsec:wsub}

We begin our testing with a bug in an isolated 
CAM output variable: \texttt{wsub}. By isolated 
we mean disconnected from the CAM core 
(see Section \ref{subsec:communities})
and highly localized. Such a bug has minimal effect and scope which 
is a good sanity check for our method. The bug consists of a 
plausible typo (transposing \texttt{0.20} to 
\texttt{2.00}) in one assignment of \texttt{wsub} 
in \texttt{microp\_aero.F90}. The variable is written to file in the next line, 
so this bug affects only the single output variable. This 
small change produces a UF-CAM-ECT failure. 
In this case the median-distance method 
clearly indicates that the wsub variable is distinct; 
the distance between the experimental and ensemble 
medians for this variable is more than 1,000 times 
greater than for the variable ranked second. 
The induced subgraph contains  
only 14 internal variables, all of which are 
related to wsub, with one being the 
bug itself. 

\subsection{RAND-MT}
\label{subsec:randmt}

\begin{figure}[t]
\begin{subfigure}[c]{0.55\textwidth}
\hspace{2ex}\includegraphics[scale=0.2]{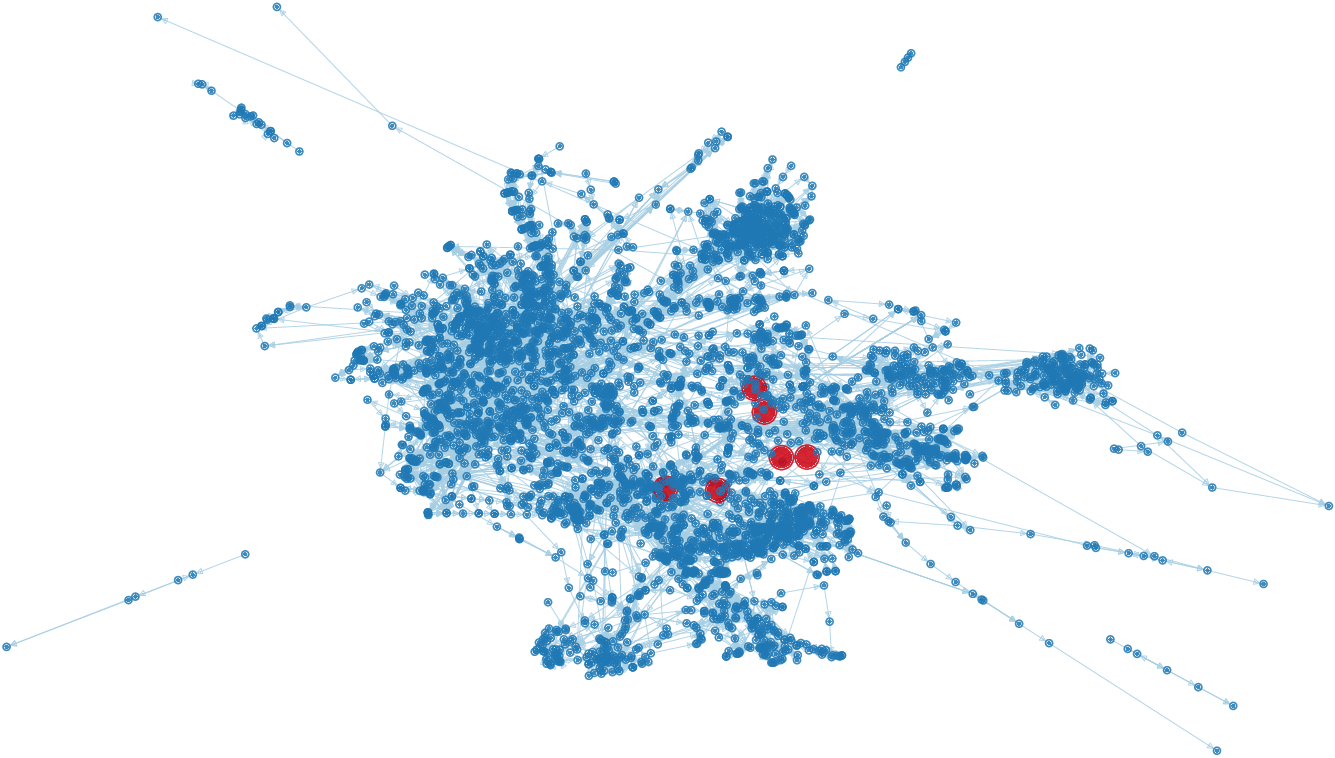}
\caption{}
\label{fig:rmt11}
\end{subfigure}

\begin{subfigure}[c]{0.55\textwidth}
\hspace{2ex}\includegraphics[scale=0.2]{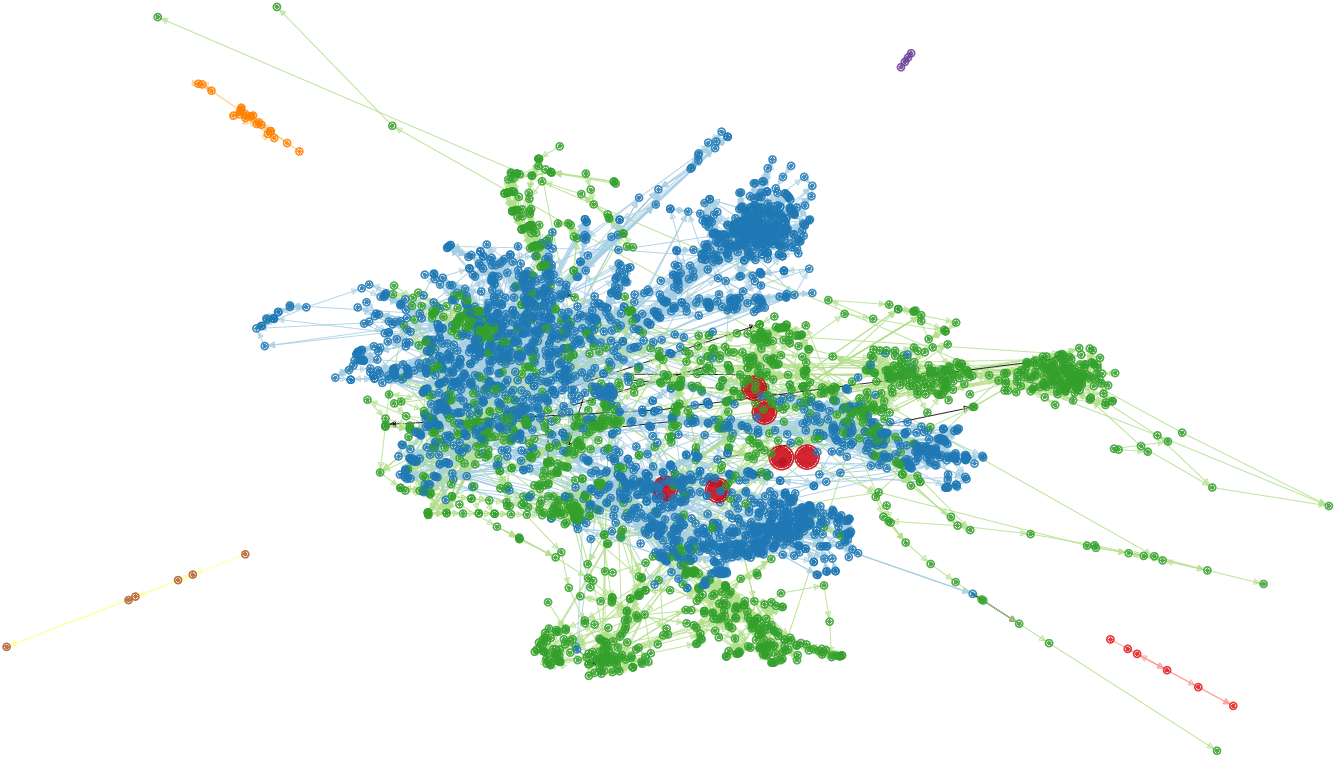}
\caption{}
\label{fig:rmt12}
\end{subfigure}

\begin{subfigure}[c]{0.55\textwidth}
\hspace{15ex}\includegraphics[scale=0.17]{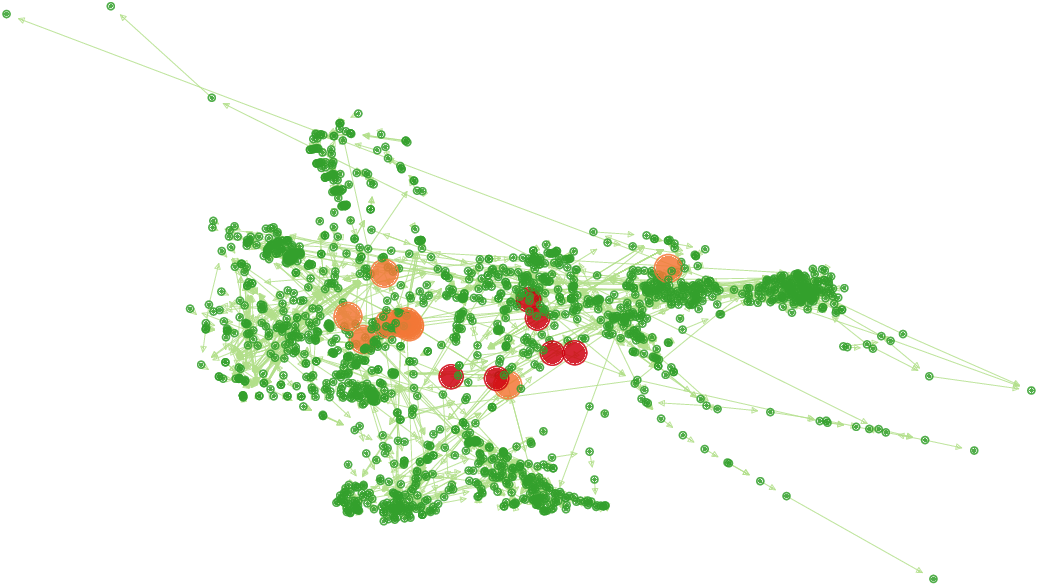}
\caption{}
\label{fig:rmt13}
\end{subfigure}
\caption{RAND-MT first iteration. Variables computed using numbers generated 
by the Mersenne Twister PRNG are larger red nodes. Subfigure \textbf{(a)} 
is the result of algorithm \ref{alg:graphrefinement} step \ref{alg:i4}, 
subfigure \textbf{(b)} colors members of each community 
detected by step \ref{alg:i5}, and subfigure \textbf{(c)} represents the output of 
step \ref{alg:i7} for the community containing 
the bugs.  Larger orange nodes in subfigure \textbf{(c)} indicate 
those with the largest eigenvector in-centrality.}
\label{fig:rmt-1}
\end{figure}

\begin{figure}[h]
\begin{subfigure}[c]{0.55\textwidth}
\hspace{2ex}\includegraphics[scale=0.12]{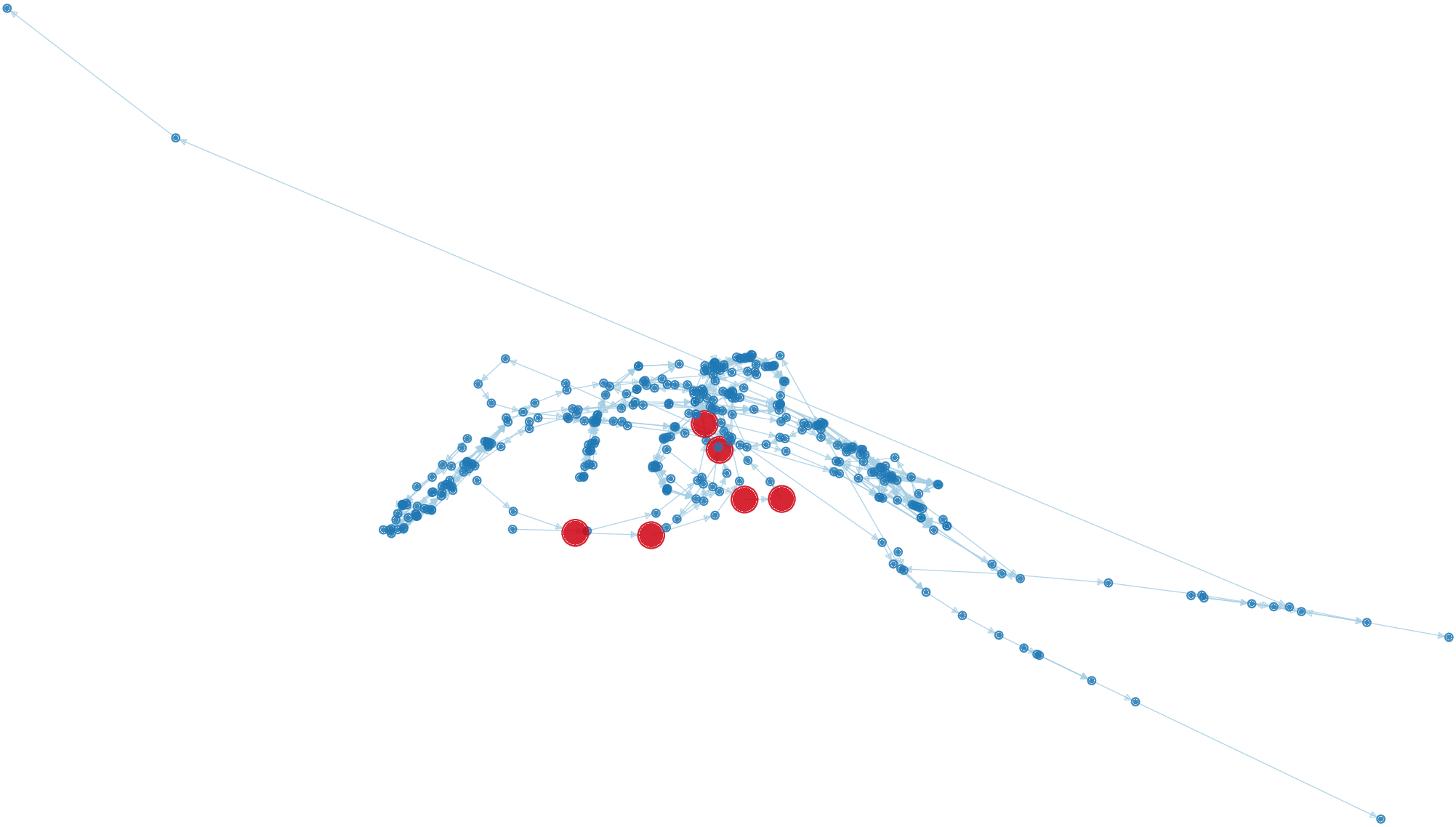}
\caption{}
\label{fig:rmt21}
\end{subfigure}

\begin{subfigure}[c]{0.55\textwidth}
\hspace{2ex}\includegraphics[scale=0.12]{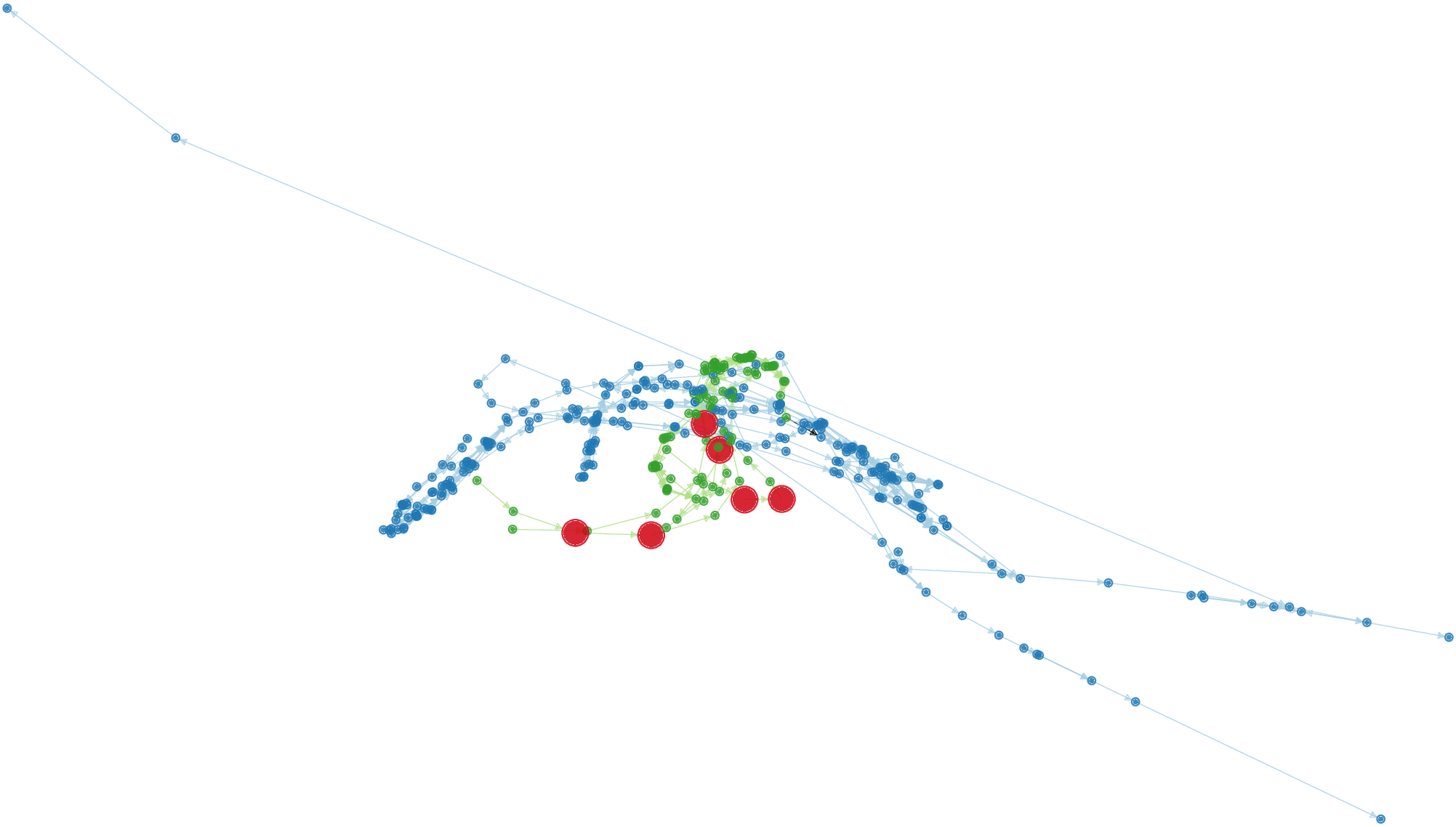}
\caption{}
\label{fig:rmt22}
\end{subfigure}

\begin{subfigure}[c]{0.55\textwidth}
\hspace{5ex}\includegraphics[scale=0.08]{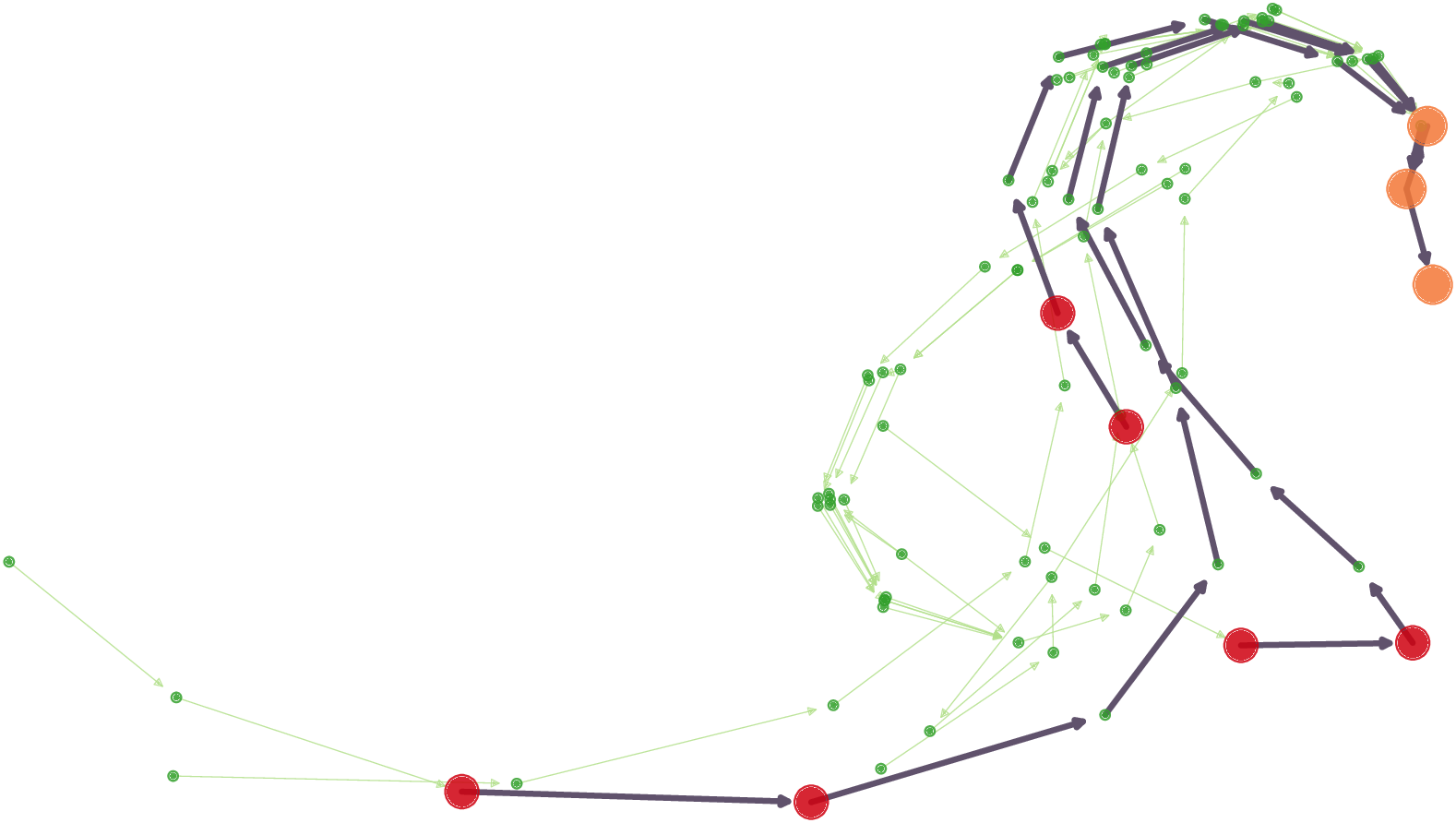}
\caption{}
\label{fig:rmt23}
\end{subfigure}

\caption[RAND-MT]
{RAND-MT second iteration. Variables computed using numbers generated 
by the Mersenne Twister PRNG are larger red nodes. 
Subfigure \textbf{(a)} is the result of 
algorithm \ref{alg:graphrefinement} step \ref{alg:i8a} 
(no different values detected), 
subfigure \textbf{(b)} colors members of each community 
detected by step \ref{alg:i5}, and subfigure 
\textbf{(c)} represents the output of 
step \ref{alg:i7} for the community containing 
the bugs.  Larger orange nodes indicate 
those with the largest eigenvector in-centrality that can be 
sampled at runtime. We choose three here given the small size of the subgraph.}
\label{fig:rmt-2}
\end{figure}

This example, RAND-MT, involves replacing the 
CESM default pseudorandom number generator (PRNG) 
with the Mersenne Twister.  This experiment 
appears in \cite{milroy2018} as an example
that results in 
a UF-CAM-ECT failure. The random number generator is used to calculate distributions
of cloud-related CAM variables, and this experiment is interesting 
because it is not a bug (in the usual sense of being 
incorrect) and not localized to a single line.
We identify the variables immediately influenced or defined by 
the numbers returned from the PRNG, and consider them to be 
the bug locations. The lasso variable selection method 
identifies the five output variables most affected by the PRNG substitution. 
From these variables, we extract a subgraph of 4,509 nodes and 9,498 edges.  
Given the size of this induced subgraph, we must use our 
iterative technique on subgraph communities 
to reduce the scope of our search. G-N identifies 
two main communities (blue and green in Fig. \ref{fig:rmt12}) in the CAM core. 
The smaller, green community contains the nodes computed using 
output from the PRNG. Instrumenting the top 10 most central variables in this community 
would not detect a difference, as there are no paths from the 
variables in the bug location to these nodes 
(see Fig. \ref{fig:rmt13}). Executing algorithm \ref{alg:graphrefinement} 
step \ref{alg:i8a} admits a dramatic reduction 
in the search space (Figure \ref{fig:rmt21}). 
Instrumenting the most central, 
orange nodes in Figure \ref{fig:rmt23} 
would indicate a difference as there are multiple paths 
from the discrepancy sources. This subgraph is small, 
and the sources are sufficiently near the sampling sites that 
the cause could be found at this stage.

It is noteworthy that the induced subgraph does 
not contain all the source locations of the 
statistical distinguishability. The PRNG in CAM is called in two 
modules: one that computes cloud cover given longwave radiation, 
and the second with shortwave radiation. The combination of 
\texttt{flwds} (downwelling longwave flux at surface) and \texttt{qrl} 
(longwave heating rate) causes the longwave module to be 
present in the induced subgraph. 
However, the two variables that are needed to include shortwave radiation 
in the induced subgraph (\texttt{fsds} and \texttt{qrs}) are not in the set 
of first five variables returned by lasso. 
The difficulty in selecting a set a variables that 
can be traced exactly to the regions of code responsible for 
statistical inconsistency means that some code sections 
can be omitted. However, the fact that many variables 
are computed in common regions of code that are 
interconnected graphically means that our method 
will identify many of these regions.

This experiment highlights an important advantage of community 
detection, namely that it separates tightly connected clusters 
and exposes smaller clusters within communities whose nodes 
dominate the centrality of the entire subgraph 
(see Section \ref{subsec:communities}).  If we were to 
sample the most central nodes of the entire subgraph (without considering communities) in 
Figure \ref{fig:rmt11}, we would be concentrating on the 
centrality-dominant blue community, and it could take many iterations of sampling 
blue community nodes to reach nodes in the green community.

\subsection{GOFFGRATCH}
\label{subsec:goffgratch}

Our third experiment is a modification in the Goff and Gratch 
Saturation Vapor Pressure elemental function. 
We change a coefficient of the water boiling temperature from \texttt{8.1328e-3} 
to \texttt{8.1828e-3}. 
This easy to miss typo results in a UF-CAM-ECT failure. 
The output of the Goff and Gratch function 
is used extensively in the CAM core, so its effects are not localized. 

\begin{figure}[t]
\begin{subfigure}[c]{0.55\textwidth}
\hspace{1ex}\includegraphics[scale=0.12]{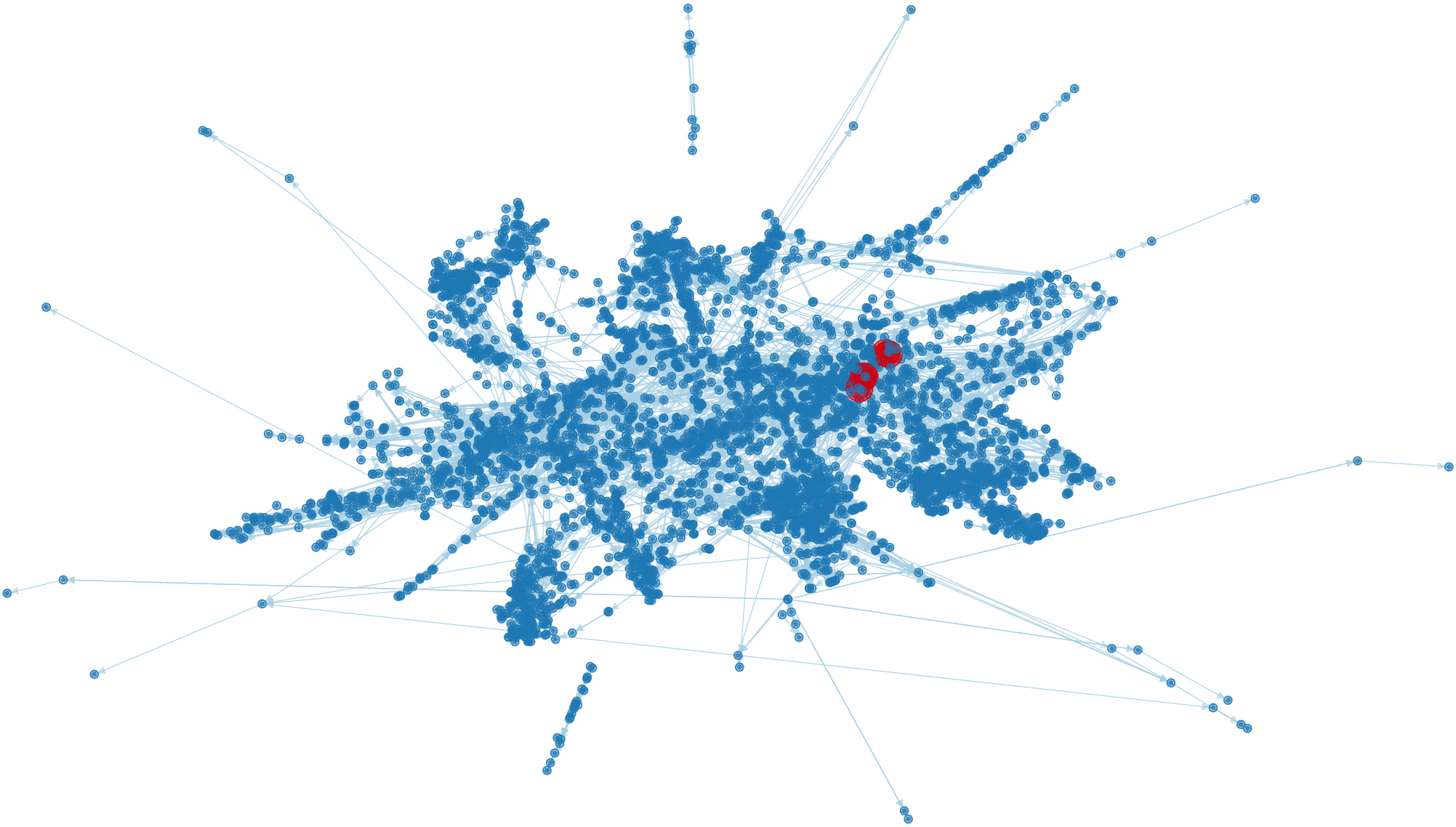}
\caption{}
\label{fig:gg11}
\end{subfigure}

\begin{subfigure}[c]{0.55\textwidth}
\hspace{1ex}\includegraphics[scale=0.12]{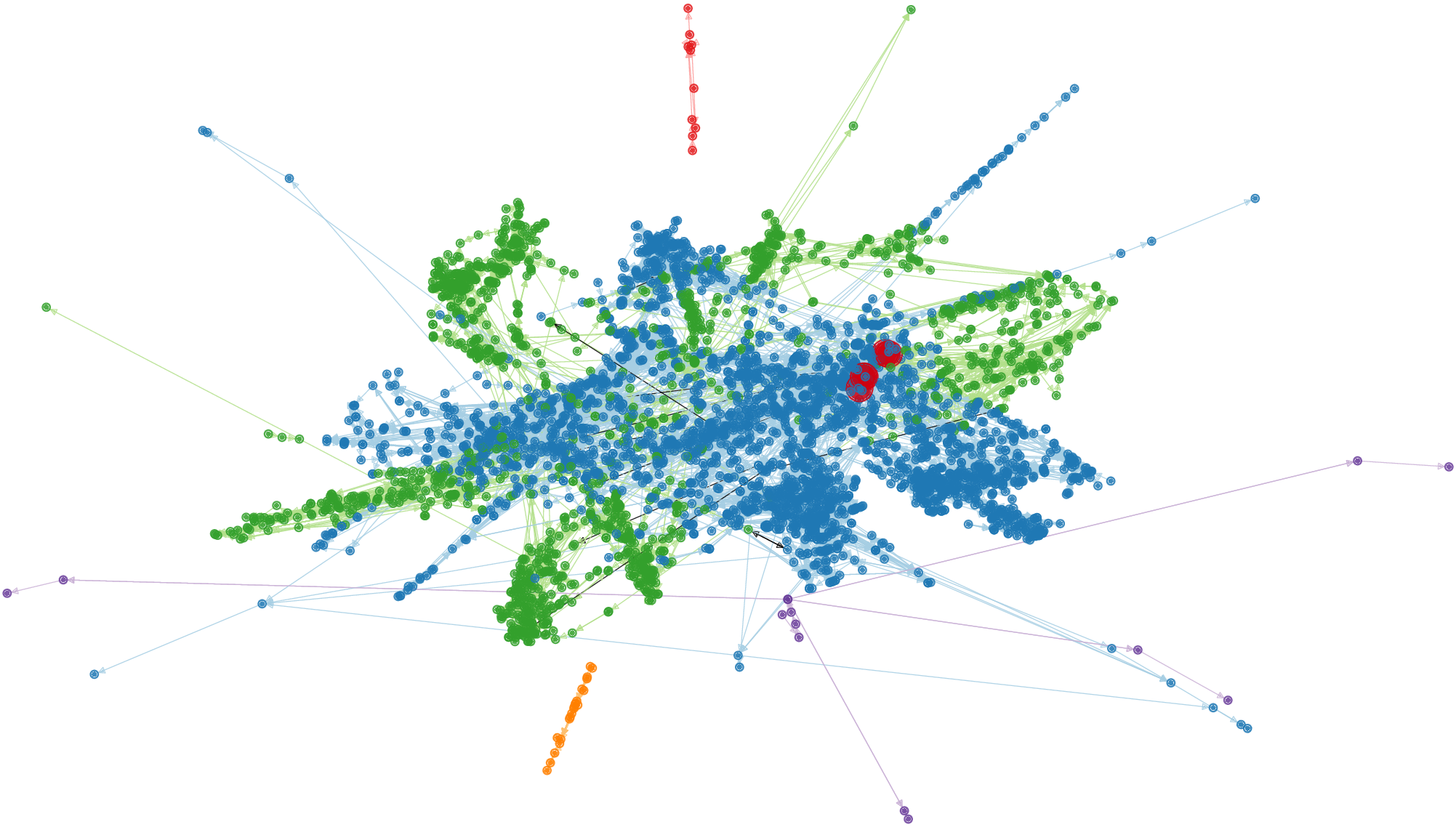}
\caption{}
\label{fig:gg12}
\end{subfigure}

\begin{subfigure}[c]{0.55\textwidth}
\hspace{5ex}\includegraphics[scale=0.12]{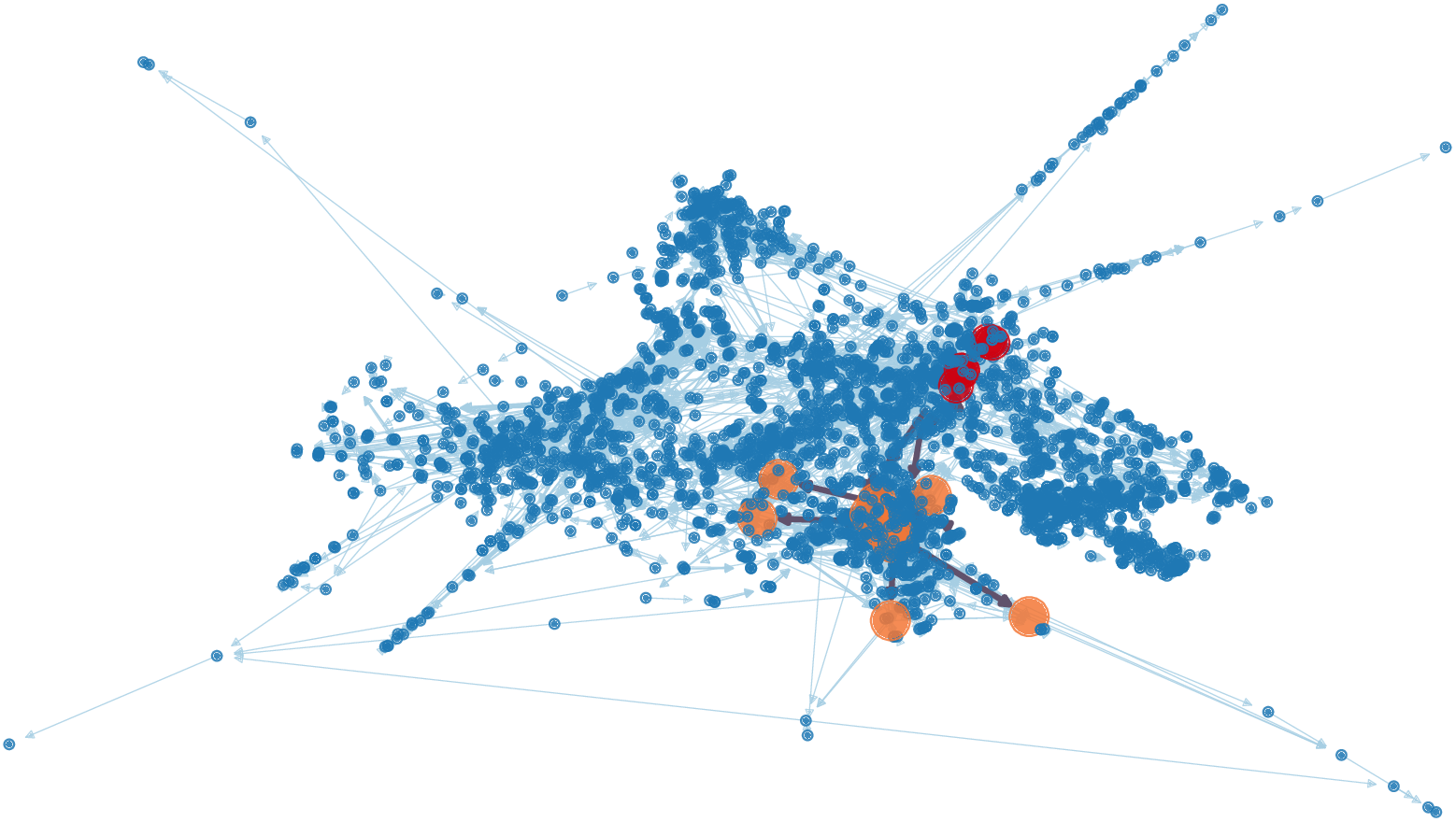}
\caption{}
\label{fig:gg13}
\end{subfigure}

\caption{GOFFGRATCH, first iteration. 
Subfigure \textbf{(a)} is the result of algorithm 
\ref{alg:graphrefinement} step \ref{alg:i4}, 
subfigure \textbf{(b)} colors members of each community 
detected by step \ref{alg:i5}, and subfigure 
\textbf{(c)} represents the output of 
step \ref{alg:i7} for the community containing 
the bugs. The bug locations are indicated as large red nodes, and the top 10 
most central variables in the blue (physics) community are indicated 
by larger orange nodes. Path segments from the bugs to the sampled 
central nodes are thicker purple edges.}
\label{fig:gg-1}
\end{figure}

For this experiment, we note that
the lasso variable selection method selects 10 variables (instead of 5). Due 
to experiment-specific conditions, tuning the 
regularization parameter to select only five variables
would require a more sophisticated approach.
Inducing a subgraph on locations that compute these 10 variables 
results in a graph of 4,243 nodes and 9,150 edges (Figure \ref{fig:gg11}). 
Note that the largest community (blue in Figure \ref{fig:gg12}) contains the  nodes affected by the incorrect coefficient. Instrumenting 
the top 10 most central variables in this community 
would detect a difference, as there are paths from the 
bug locations to these nodes 
(Figure \ref{fig:gg13}). 

For the second iteration, inducing a new 
subgraph of all shortest paths terminating on the 
central nodes in Figure \ref{fig:gg13} (algorithm \ref{alg:graphrefinement} 
step \ref{alg:i8b}) returns a subgraph that includes 
part of the green community from the first iteration.
Subsequent community detection reveals the remnants 
of the green community of the first iteration, 
which are then excluded by sampling. However, in 
this case, no further simulated iterative 
refinement can be performed by inducing 
a subgraph on nodes connected to the 
instrumented variables, as this subgraph is 
so highly connected that the induced subgraph equals 
the community subgraph (nearly 
identical to Figure \ref{fig:gg13}). We omit 
plots of the second iteration due to  
similarity with Figure \ref{fig:gg-1}.

To refine the blue community from Figure \ref{fig:gg13} 
we need to handle the case where the induced 
subgraph does not refine the subgraph from the 
previous iteration. In future work we can 
rank the differences obtained by sampling and 
further refine the subgraph based on the 
nodes with the greatest differences. 
Alternatively, if we learn that all nodes 
are affected equally, we can choose one node 
and induce a subgraph based on paths terminating on it. 
There are numerous algorithms for graph partitioning 
which we could use in conjunction with sampling 
in the case of GOFFGRATCH, and there are many options to 
test when we realize our process with sampled data
(which is also future work). The power of our graphical method 
is that it enables diverse types of graph analysis. 

\subsection{AVX2}
\label{subsec:AVX2}

As noted in Section \ref{sec:intro}, our work here to develop an automated process was motivated by our
investigation in \cite{milroy2016} to find the cause of the CESM-ECT failure on the Mira 
supercomputer \cite{mira2018} when compared to the ensemble from NCAR's Yellowstone machine \cite{yellowstone2016}. 
To summarize briefly, the manual process in \cite{milroy2016} included measuring each CAM output variable's 
contribution to the CAM-ECT failure rate, which  
identified the most affected output variables. We were then able to determine that computations 
in the  Morrison-Gettelman microphysics module (MG1) were problematic \cite{milroy2016}, and, with the previously mentioned KGen tool
\cite{kim2016,kim2017}, convert the MG1 module into a kernel.  In the MG1 kernel, we found variables 
which had substantially different values between Yellowstone and 
Mira and were able to attribute the difference to  Mira's FMA instructions.
We note (for later comparison) that one of these variables was \texttt{nctend}, which is 
modified by a frequently used temporary variable \texttt{dum} and that 
\texttt{nctend} also exhibits significantly different values between Yellowstone
and Haswell generation (FMA capable) Intel CPUs (i.e., the FMA discrepancy is not limited to Mira).
In this section, we demonstrate our automated method could obtain these same results that took several months and many CESM experts.

Because we are no longer able to use Mira and Yellowstone, 
we instead evaluate the impact of FMA on the newer Cheyenne machine \cite{cheyenne2017}. Cheyenne 
contains Intel Broadwell CPUs, which support the Intel AVX2 
instruction set, and these instructions include FMA.
For this work, we compare an ensemble generated with AVX2 disabled 
(thus disabling FMA) to an experimental set generated with AVX2 
(and FMA) instructions enabled. We verify that enabling AVX2 and FMA causes 
a UF-CAM-ECT failure (see Table \ref{tab:avx2}). Since FMA instructions can be 
generated from many different lines of source code (distributed 
sources of discrepancy), we employ KGen to identify a small 
number of variables affected by AVX2 and FMA to designate 
as bugs. We extract the Morrison-Gettelman microphysics kernel 
identified in \cite{milroy2016} and compare the normalized Root 
Mean Squared (RMS) values computed by the kernel with AVX2 
disabled to the normalized RMS values with AVX2 enabled. 
KGen flags 42 variables as exhibiting  
normalized RMS value differences exceeding $10^{-12}$. Here, we
determine if our iterative refinement procedure 
can find some of these variables given CAM outputs 
most affected by AVX2 instructions.

\begin{figure}[t]
\begin{subfigure}[c]{0.55\textwidth}
\hspace{1ex}\includegraphics[scale=0.17]{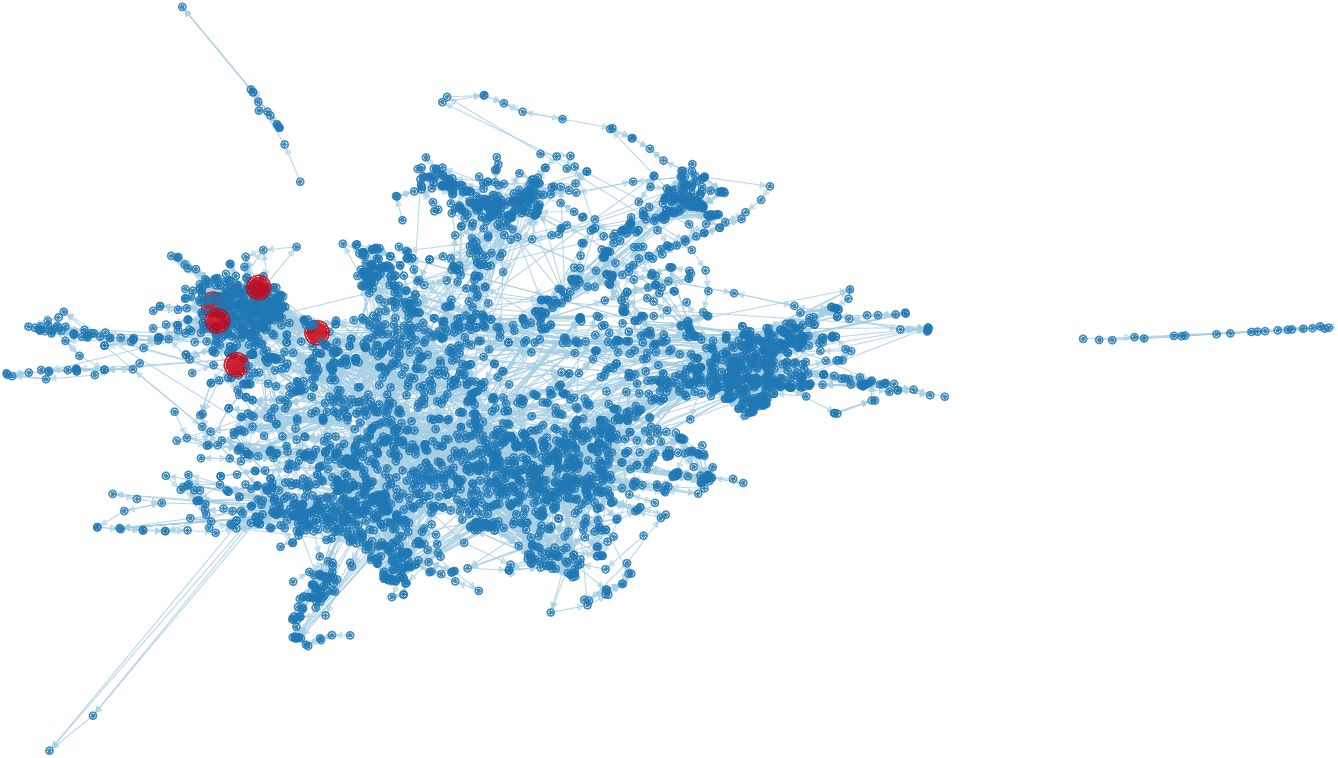}
\caption{}
\label{fig:avx21}
\end{subfigure}

\begin{subfigure}[c]{0.55\textwidth}
\hspace{1ex}\includegraphics[scale=0.17]{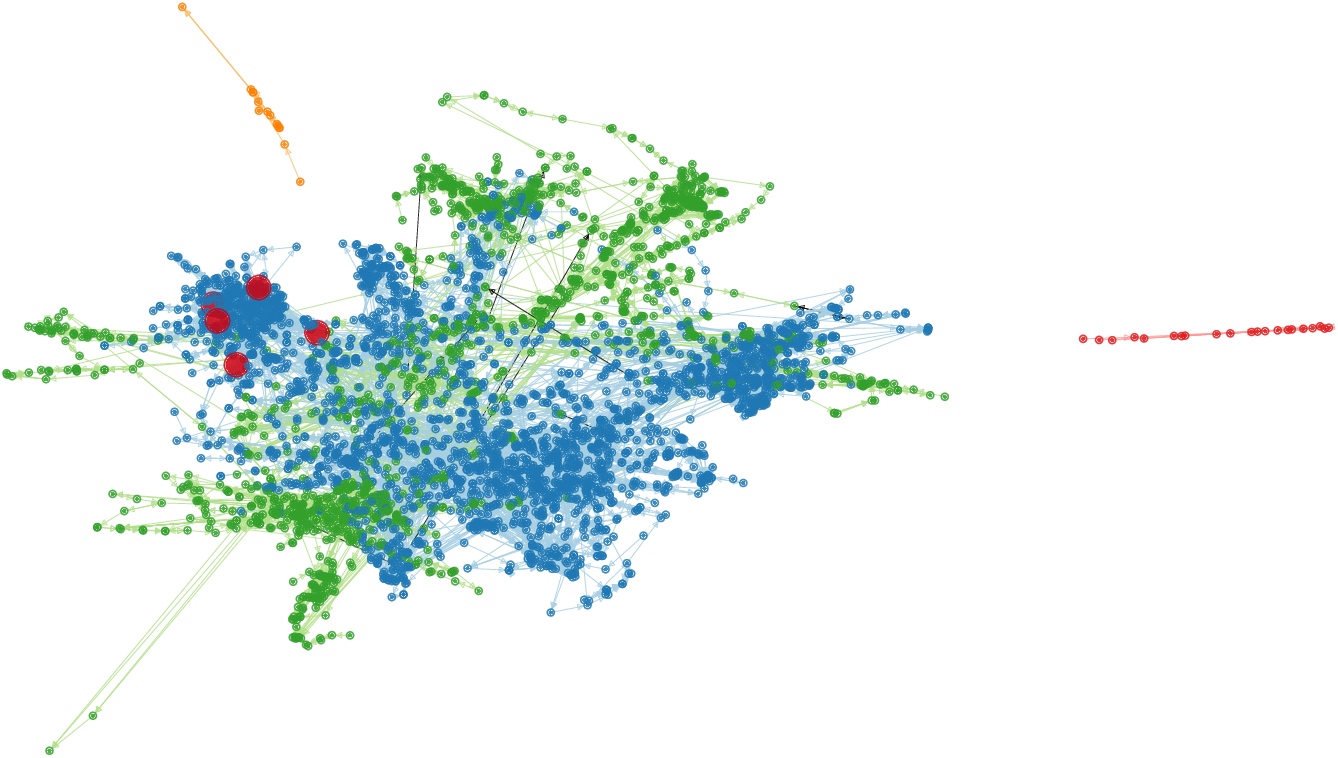}
\caption{}
\label{fig:avx22}
\end{subfigure}

\begin{subfigure}[c]{0.55\textwidth}
\hspace{5ex}\includegraphics[scale=0.14]{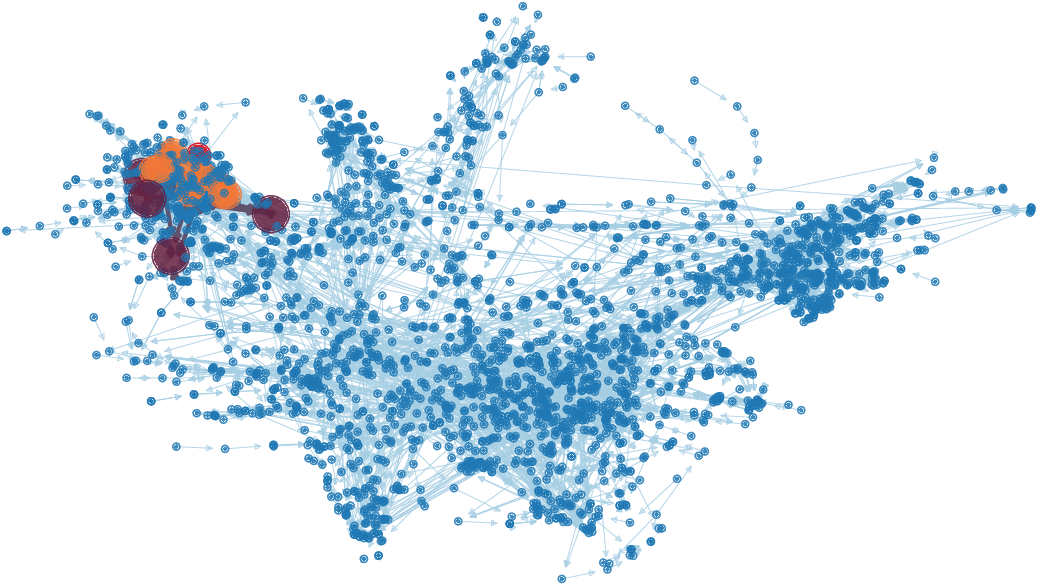}
\caption{}
\label{fig:avx23}
\end{subfigure}

\caption[AVX2]{AVX2. Variables found to take significantly different 
normalized RMS values between Broadwell CPUs with AVX2 enabled (FMA enabled) 
and AVX2 disabled (FMA disabled) are larger red nodes 
(\texttt{nctend}, \texttt{qvlat}, \texttt{tlat}, \texttt{nitend}, and 
\texttt{qsout} in MG1) and also in the top 15 most central nodes. 
Large orange nodes are remaining nodes in the top 15. 
Subfigure \textbf{(a)} is the result of algorithm 
\ref{alg:graphrefinement} step \ref{alg:i4}, 
subfigure \textbf{(b)} colors members of each community 
detected by step \ref{alg:i5}, and subfigure 
\textbf{(c)} represents the output of 
step \ref{alg:i7} for the community containing 
the bugs. Bugs detected by sampling are colored purple.}
\label{fig:avx2}
\end{figure}

Inducing a subgraph on assignment paths that compute CAM output variables 
affected by enabling AVX2 instructions (selected by lasso) results in the graph 
in Figure \ref{fig:avx21} (4,159 nodes and 9,028 edges). 
Five of the 42 variables identified by KGen are present in this 
subgraph, all of which are in the blue community of Figure \ref{fig:avx22}. 
This community contains the CAM core physics processes, 
of which MG1 forms a central part. The node with the largest 
eigenvector in-centrality is the temporary, dummy 
variable \texttt{dum} in Figure \ref{fig:avx23}.  
Four of the five variables with normalized RMS values exceeding 
our threshold are in the top 15 nodes with the greatest in-centrality. These variables are 
\texttt{nctend}, \texttt{qvlat}, \texttt{tlat}, and 
\texttt{nitend}. The fifth variable, (\texttt{qsout}), 
is modified by \texttt{qniic} (in the top 15 most central 
nodes) in an assignment statement. All five variables have 
paths that terminate on all 15 most central nodes. 
The following Python REPL output lists the nodes in the blue 
community of Figure \ref{fig:avx23} in descending order of their 
eigenvector in-centrality values. Each ordered pair contains 
a node name and a centrality value. The node name includes a 
suffix demarcated by a double underscore; this suffix 
indicates the subprogram containing the variable (here the \texttt{micro\_mg\_tend}
subroutine in the Morrison-Gettelman microphysics kernel) to guarantee 
unique names in the directed graph (see Section \ref{subsec:ast-digraph}). 
Node names colored red are those exhibiting values exceeding 
our normalized RMS threshold.

\noindent \small \texttt{>>> avx2\_bluecommunity\_incentrality[:16]} \\
\noindent \footnotesize \texttt{(dum\_\_micro\_mg\_tend, 0.455153), (ratio\_\_micro\_mg\_tend, 0.325264), \newline
(\textcolor{red}{tlat\_\_micro\_mg\_tend}, 0.255383), (qniic\_\_micro\_mg\_tend, 0.198578), \newline
(nric\_\_micro\_mg\_tend, 0.196431), (nsic\_\_micro\_mg\_tend, 0.191075), \newline
(qctend\_\_micro\_mg\_tend, 0.188477), (qric\_\_micro\_mg\_tend, 0.180318), \newline
(qitend\_\_micro\_mg\_tend, 0.15969), (prds\_\_micro\_mg\_tend, 0.157626), \newline
(pre\_\_micro\_mg\_tend, 0.157551), (\textcolor{red}{nctend\_\_micro\_mg\_tend}, 0.148088), \newline
(\textcolor{red}{qvlat\_\_micro\_mg\_tend}, 0.132584), (mnuccc\_\_micro\_mg\_tend, 0.121525), \newline
(\textcolor{red}{nitend\_\_micro\_mg\_tend}, 0.120172), (nsagg\_\_micro\_mg\_tend, 0.109382)} \normalsize

\noindent That our iterative refinement procedure would 
sample and identify the locations of nodes 
known to be most affected by AVX2 instructions 
on the first iteration is a testament to the 
potential utility of our method, particularly in the 
challenging case where hardware or CPU instructions
cause statistical distinguishability.

\subsection{AVX2 in the CESM graph}
\label{subsec:avx2-cesm-graph}

Here, we deviate slightly to discuss how centrality can be used to 
identify Fortran modules crucial to information flow in the overall CESM graph.
While the MG1 module and its constituent variables are causes of 
ECT failure with AVX2 and FMA enabled, these instructions can be generated in many CESM modules. 
This suggests we compute the (in and out) centrality of the modules themselves 
(rather than individual variables) to rank them by 
their potential to propagate FMA-caused differences within CESM. This viewpoint 
applies to other machine instructions or hardware errors. 

To calculate the centrality, we must collapse the graph of variables into 
modules by considering the graph minor of CESM code formed by the quotient 
graph of Fortran modules. A graph minor is a subgraph of a graph $G$ 
obtained by contracting edges of $G$. This graph minor identifies 
(or collapses) nodes using an equivalence relation, meaning 
that if two nodes satisfy the equivalence relation, they are replaced by a 
single node.  Edges between equivalent nodes are deleted, and edges between 
remaining nodes and the new node are preserved.  In this case we use 
the equivalence relation $v_1 \thicksim v_2 \iff v_1$ and $v_2$ are in the same 
CESM module (modules become equivalence classes). 
Applying this equivalence relation to the CESM graph yields 
a digraph of 561 nodes and 4,245 edges. Selectively disabling 
AVX2 on the top 50 modules ranked by 
centrality results in a substantial 
reduction in the UF-CAM-ECT failure rate in comparison with AVX2 
enabled on all modules. Furthermore, this approach exhibits a substantially lower 
failure rate than disabling AVX2 on 50 modules at random, and even 
the top 50 modules by lines of code. See Table \ref{tab:avx2} 
for the failure 
rates. These results indicate that eigenvector centrality
accurately captures the information flow between CESM modules and provides a 
useful ordering. Selective disablement of instructions such as 
AVX2 balances optimization with preserving statistical consistency 
and promotes highly efficient CPU usage. 

\begin{table}
  \caption{Selective AVX2 disablement}
  \label{tab:avx2}
  \begin{tabularx}{0.48\textwidth}{lc}
    \toprule
    Experiment & ECT failure \\
               & rate \\
    \midrule
    AVX2 enabled, all modules & 92\%  \\
    AVX2 disabled, 50 largest modules & 86\% \\
    AVX2 disabled, 50 rand mods (10 sample avg)& 83\% \\
    AVX2 disabled, 50 central modules & 8\% \\
    AVX2 disabled, all modules & 2\% \\
  \bottomrule
\end{tabularx}
\end{table}

\section{Conclusions and future work}
\label{sec:conclusion}

The goal of this study is to develop 
methods that make root cause analysis of CESM 
possible. To this end, 
we create a toolkit to convert the CESM source 
code into a digraph together with metadata 
that represents variable assignment paths and 
their properties. We note that creating a Python interface for LLVM \cite{lattner2004} with Flang 
\cite{flang2017} would allow our parsing to succeed on 
any compilable Fortran code. We develop an efficient 
hybrid static program slicing approach 
based on combining code coverage with BFS. 
We combine the Girvan-Newman algorithm with 
eigenvector in-centrality in series to 
enable parallel runtime sampling of critical nodes. 
We perform experiments based on CESM output to demonstrate 
in simulation how our process can find causes of model 
discrepancy. Finally, we provide evidence that our methods 
accurately characterize information flow at runtime by 
successfully finding variables determined to be 
susceptible to FMA instructions by a lengthy, manual investigation.
Creating a method to identify which variables to sample 
to refine the root cause search space is a significant 
accomplishment.  However, developing and implementing a sampling procedure 
for the running model is a challenging undertaking that 
remains to be done. 

The power of the methods developed in this work is derived from 
the translation of source code into a directed graph representation 
that enables fast, static analysis of information flow. Graph theory is an 
expansive field that offers a large number of sophisticated ways 
to analyze variable relationships. Allowing scientists and engineers 
to apply these techniques to root cause analysis is a significant 
achievement. In future work, combining the static directed graph with runtime 
values into a fully-featured root cause analysis suite for CESM will 
provide a level of quality assurance not currently available. 
Modifying our method to enable quality assurance for other codes is future work.

\section*{Acknowledgements}
We wish to thank Dong Ahn, John Dennis, and Sriram Sankaranarayanan for their helpful advice. 
We are especially grateful to Ganesh Gopalakrishnan, Michael Bentley, and Ian Briggs for 
thoroughly reading our draft and for their detailed feedback.
This research used computing resources provided by the Climate Simulation Laboratory at 
NCAR's Computational and Information Systems Laboratory (CISL), sponsored by the 
National Science Foundation and other agencies.

\printbibliography

\clearpage


\section{Supplementary material}

\subsection{Hashimoto non-backtracking centrality}
\label{supp:nbtcentrality}

\begin{figure}[t]
\includegraphics[width=0.48\textwidth]{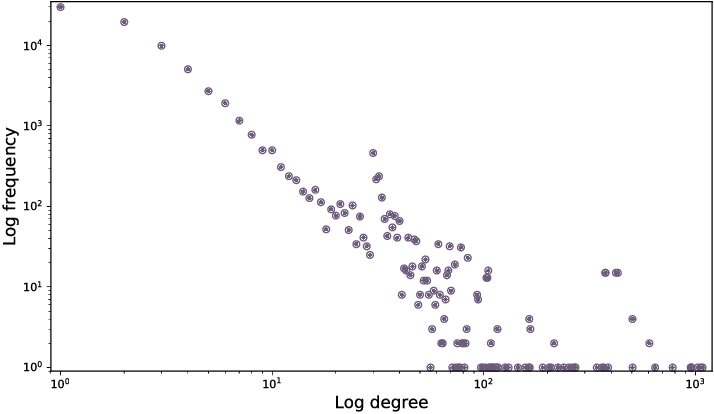}
\caption[CESM degree dist]
{Degree distribution of nodes in the CESM digraph.}
\label{fig:supp-cesm-deg-dist}
\end{figure}

\begin{figure}[t]
\includegraphics[width=0.48\textwidth]{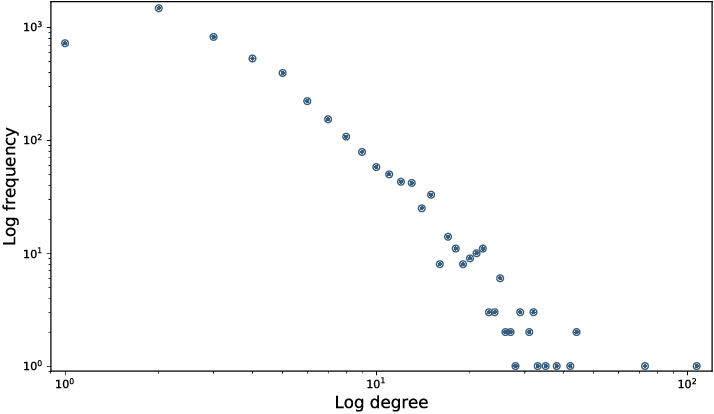}
\caption[GoffGratch degree dist]
{Degree distribution of nodes in the GOFFGRATCH digraph.}
\label{fig:supp-gg-deg-dist}
\end{figure}

\begin{figure}[t]
\includegraphics[width=0.48\textwidth]{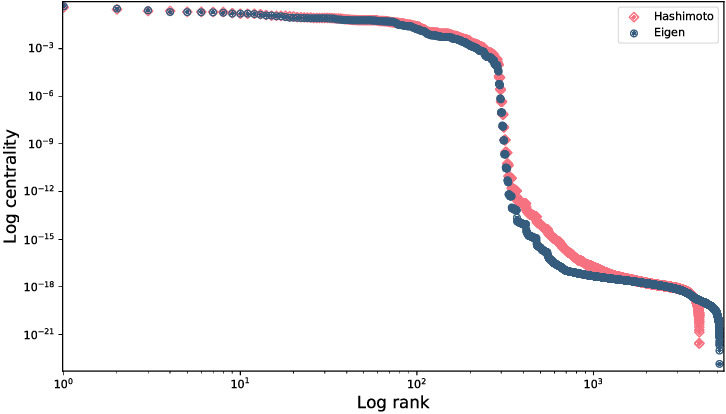}
\caption[GoffGratch: eigen vs. hashi]
{Log rank versus log absolute value of centrality for Hashimoto non-backtracking centrality 
and eigenvector centrality in the GOFFGRATCH experiment subgraph. The 
absolute value of the centralities is used since the lowest ranked terms 
are small negative numbers.}
\label{fig:supp-gg-hashi-eigen}
\end{figure}

Scale-free or power law graphs which have 
degree distributions that are negative exponentials 
with exponent magnitude greater than 2.5 are identified as
causing localization in \citep{martin2014}. 
The degree distribution of the total CESM graph 
approximately follows a power law, as can be seen in 
Figure \ref{fig:supp-cesm-deg-dist}. Induced subgraphs 
of the CESM graph are also approximately 
scale-free, consistent with the properties 
of such graphs (see Figure \ref{fig:supp-gg-deg-dist} 
for the GOFFGRATCH experiment subgraph). 
A natural question is whether the concentration 
of centrality on graph hubs has undesirable 
effects on the ranking of nodes. 
The application of non-backtracking or Hashimoto 
centrality \cite{hashimoto1989} as 
a substitute for eigenvector centrality 
for power law graphs is discussed in \cite{martin2014}. We compare the two 
centralities in Figure \ref{fig:supp-gg-hashi-eigen} 
for the GOFFGRATCH experiment. The Hashimoto 
non-backtracking centrality indeed distributes 
the centrality from the hubs to other nodes, 
but the effect is subtle until approximately 
the $300^{th}$ ranked node. Also note that the 
Hashimoto centrality does not provide a 
rank for all nodes in the subgraph, as can 
be noted by the sharp drop at the end 
of its curve in Figure \ref{fig:supp-gg-hashi-eigen}. 
This is due to the Hashimoto centrality's 
use of the line graph of the subgraph's 
adjacency matrix, which excludes nodes 
with no neighbors. Although we determine 
that the non-backtracking centrality provides 
at best marginal improvement over eigenvector 
centrality for our graph, we provide a 
derivation based on that which appears 
in \citep{martin2014}. Hashimoto 
centrality may prove beneficial for 
models with graphs that follow power 
laws that produce more pronounced localization 
\citep{martin2014}.

\subsubsection{Centrality derivation}
\label{supp:nbt-derived}
This section is a reformulation of the derivation in \cite{martin2014}, which we 
have reworked in the interest of clarity.
Let $G$ be a graph with $n \times n$ adjacency matrix $\mathbf{A}$, set of nodes $V$ and edges $E$. 
The graph order is the number of nodes: $|V| = n$, while the graph size is the number of 
edges: $|E| = m$. Note that $m$ is the number of nonzero entries in $\mathbf{A}$ if $G$ is 
directed, and the number of nonzero entries in the upper or lower triangle of $\mathbf{A}$ 
if $G$ is undirected. 
\medskip \\
Let $e \in E$ be represented as $(v_1, v_2)$.  If $G$ is directed, $(v_1, v_2) \neq (v_2, v_1)$ 
and the order represents direction: $v_1 \rightarrow v_2 \coloneqq (v_1, v_2)$.  If 
$G$ is undirected, $(v_1, v_2) = (v_2, v_1)$.
\medskip \\
Let $N(i)$ be the set of neighbors of node $i$. \\

The Hashimoto, or non-backtracking matrix \cite{hashimoto1989} of graph $G$ is denoted $\mathbf{B}$ and is an adjacency matrix on $E$:

\begin{align*}
& \forall (u, v), (w, x) \in E; \\
B_{(u, v), (w, x)} \;& \text{or} \; B_{(u \rightarrow v), (w \rightarrow x)} = \delta_{vw} (1 - \delta_{ux})
\end{align*}

Where $\delta$ is the Kronecker delta.  For an undirected graph, each $(v_1, v_2) \in E$ 
becomes two ordered pairs $(v_1, v_2), (v_2, v_1)$. Thus $\mathbf{B}$ is 
$m \times m$ if $G$ is directed, and $2m \times 2m$ for undirected $G$.  $\mathbf{B}$ 
is closely related to the \textit{line graph} $L(G)$ which is also an adjacency matrix on 
$E$: $L(G)_{(u \rightarrow v), (w \rightarrow x)} \coloneqq \delta_{vw}$ \\

Instead of computing the eigenvector centrality on $\mathbf{A}$, we use $\mathbf{B}$.  
Let $\lambda$ be the Perron-Frobenius (leading) eigenvalue of $\mathbf{B}$, 
and $\vec{v}$ be the corresponding eigenvector.  Then the out-centrality 
(corresponding to out-edges) for some $i \in V$ can be derived by 
starting from the eigenvector equation $\lambda \vec{v} = \mathbf{B}\vec{v}$. 
To compute the in-centrality used in this work, we can reverse 
the directed edges of $\mathbf{A}$ (via the transpose $\mathbf{A}^\intercal$).

\begin{align*}
\vec{v}_{(i\rightarrow j)} &= \frac{1}{\lambda} \sum_{(k \rightarrow l) \in N((i \rightarrow j))} B_{(i \rightarrow j), (k \rightarrow l)} \vec{v}_{(k \rightarrow l)} \\
&= \frac{1}{\lambda} \sum_{(k \rightarrow l) \in N((i \rightarrow j))} \delta_{jk} (1 - \delta_{il}) \vec{v}_{(k \rightarrow l)}\\
&= \frac{1}{\lambda} \sum_{k}^n \sum_l^n A_{kl} \delta_{jk} (1 - \delta_{il}) \vec{v}_{(k \rightarrow l)} \\
&= \frac{1}{\lambda} \sum_l^n A_{jl} (1 - \delta_{il}) \vec{v}_{(j \rightarrow l)} \\
\text{or} \\
\vec{v}_{(i\rightarrow j)} &= \frac{1}{\lambda} \sum_{l \neq i}^n A_{jl} \vec{v}_{(j \rightarrow l)} \\
\intertext{then the full non-backtracking centrality of node  $i$  is:} \\
c_i &= \sum_{q \in N(i)} \vec{v}_{(i\rightarrow q)} \\
\end{align*}

Where we are free to choose a constant to normalize the centrality.

\subsection{Additional experimental results}
\label{supp:experiments}
As in Section \ref{sec:experiments} we omit communities of 
fewer than four nodes in the interest of plot clarity. 
Note that we refer to the Girvan-Newman algorithm (\cite{girvan2002,newman2004}) as 
G-N. 

\begin{table*}[t]
  \caption{CAM output variables selected by the methods described in paper Section \ref{sec:varselect}, 
  and their internal counterparts.}
  \label{tab:supp-varselec}
  \begin{tabularx}{\textwidth}{lXX}
    \toprule
    Experiment & Output variables & Internal variables \\
    \midrule
    WSUBBUG & wsub & wsub \\
    RANDOMBUG & omega & omega \\
    GOFFGRATCH & aqsnow, freqs, cldhgh, precsl, ansnow, cldmed, cloud, cldlow, ccn3, cldtot & qsout2, freqs, clhgh, snowl, nsout2, clmed, cld, cllow, ccn, cltot \\
    DYN3BUG & vv, omega, z3, uu, omegat & v, omega, z3, u, t \\
    RAND-MT & flds, taux, snowhlnd, flns, qrl & flwds, wsx, snowhland, flns, qrl \\
    AVX2 & taux, trefht, snowhlnd, ps, u10, shflx & wsx, tref, snowhland, ps, u10, shf \\
  \bottomrule
\end{tabularx}
\end{table*}

\subsubsection{RANDOMBUG}

\begin{figure*}[th!]
\begin{subfigure}[c]{\textwidth}
\hspace{15ex}\includegraphics[scale=0.37]{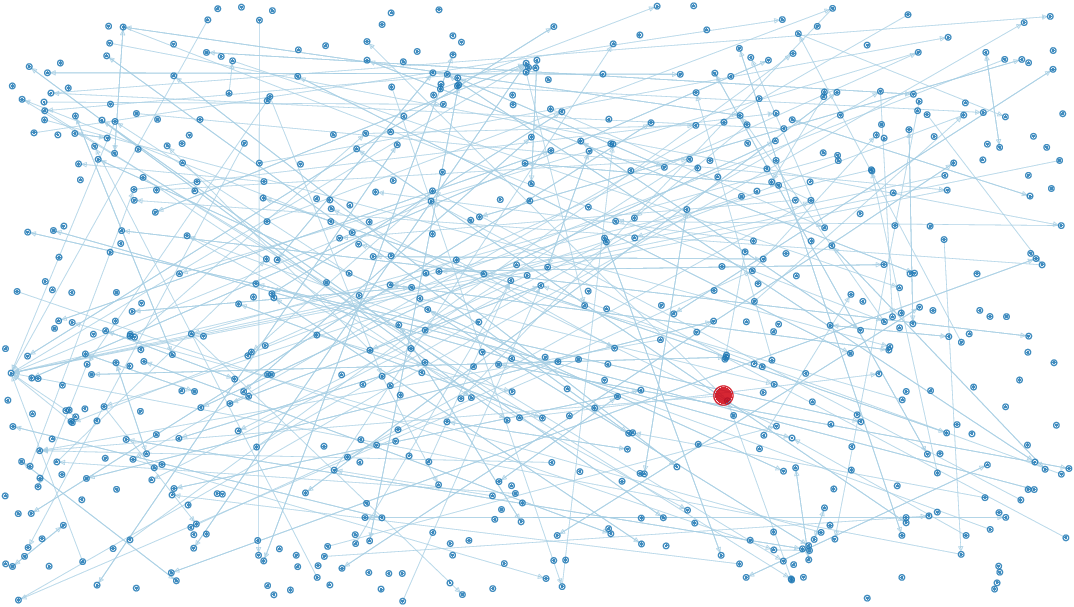}
\caption{}
\label{fig:supp-randombug1}
\end{subfigure}

\begin{subfigure}[c]{\textwidth}
\hspace{15ex}\includegraphics[scale=0.37]{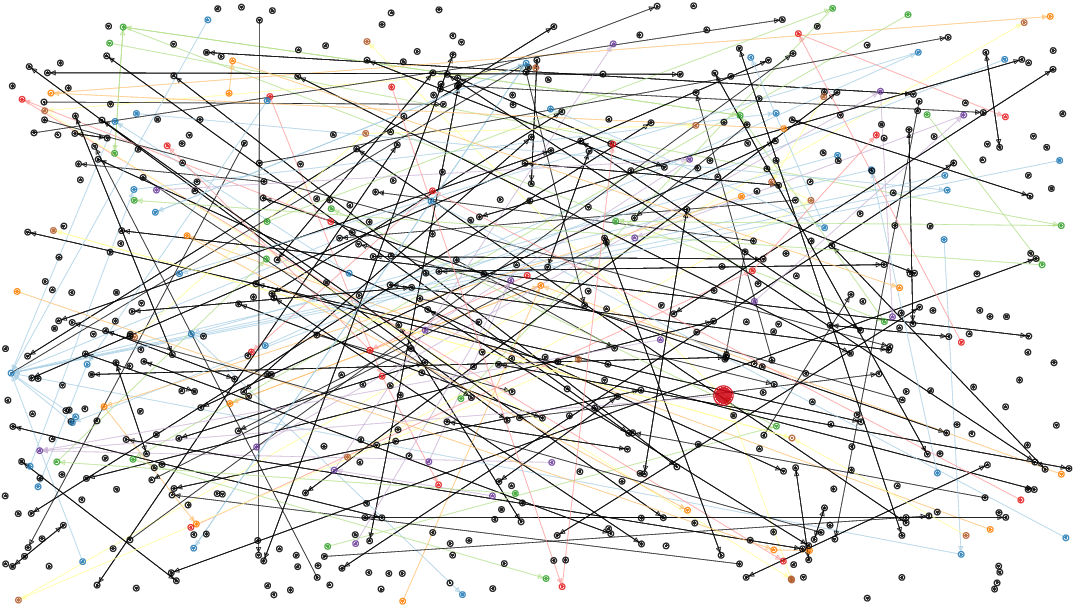}
\caption{}
\label{fig:supp-randombug2}
\end{subfigure}

\begin{subfigure}[c]{\textwidth}
\hspace{24ex}\includegraphics[scale=0.27]{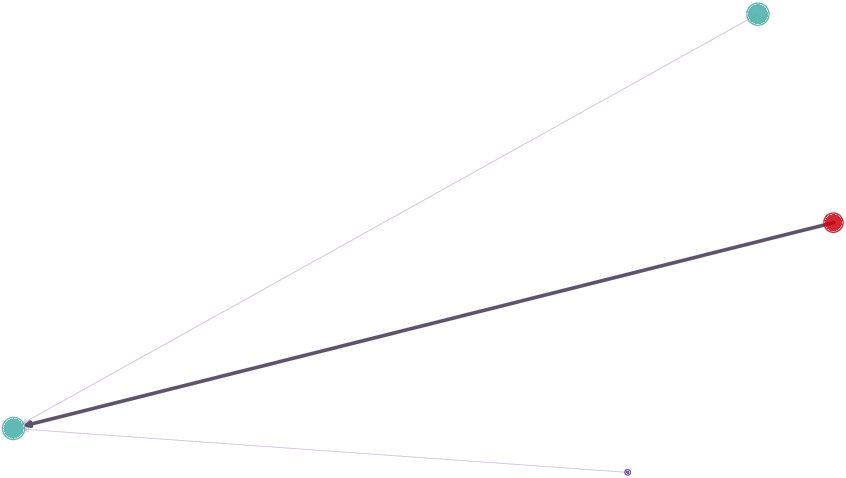}
\caption{}
\label{fig:supp-randombug3}
\end{subfigure}

\caption{RANDOMBUG, single iteration. The bug location is indicated 
by a large red node. In \textbf{c}, the light blue nodes are the most central 
of the small community shown, and the purple edge designates the connection 
from the bug to the instrumented node.}
\label{fig:supp-randombug}
\end{figure*}

We select the module for this bug by randomly choosing 
a module from the set of CAM modules known to be executed by our 
simulation in the first time step.  We introduce an error in the 
array index of a variable used to assign the contents of the 
derived type containing physics state variables (t, u, v, etc.), 
in particular the state variable omega. As in the previous experiment, 
this change results in a UF-CAM-ECT failure. Omega is output to 
file with the value \texttt{state\%omega}, so we use 
``omega'' as the canonical name for generating the 
induced subgraph. This experiment is more challenging 
than WSUBBUG, as omega is computed in other CAM modules, 
yielding a subgraph of 628 nodes and 295 edges.  
Applying the G-N algorithm to the remaining nodes identifies
several small (fewer than 30 nodes) communities, one of 
whose most central node is the bug source. See Figure \ref{fig:supp-randombug}.

\subsubsection{DYN3BUG}
\label{subsec:supp-dyn3bug}

\begin{figure*}[th!]
\begin{subfigure}[c]{\textwidth}
\hspace{15ex}\includegraphics[scale=0.27]{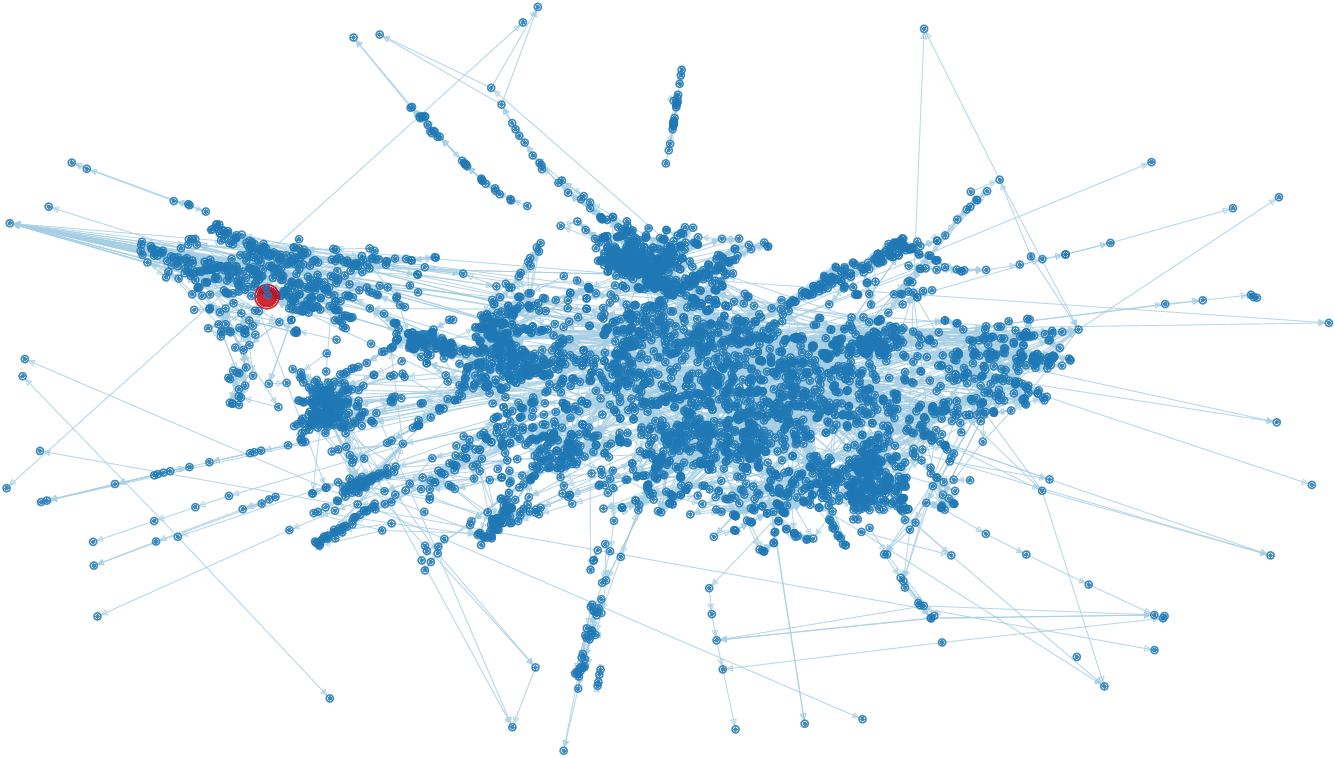}
\caption{}
\label{fig:supp-dyn3b11}
\end{subfigure}

\begin{subfigure}[c]{\textwidth}
\hspace{15ex}\includegraphics[scale=0.27]{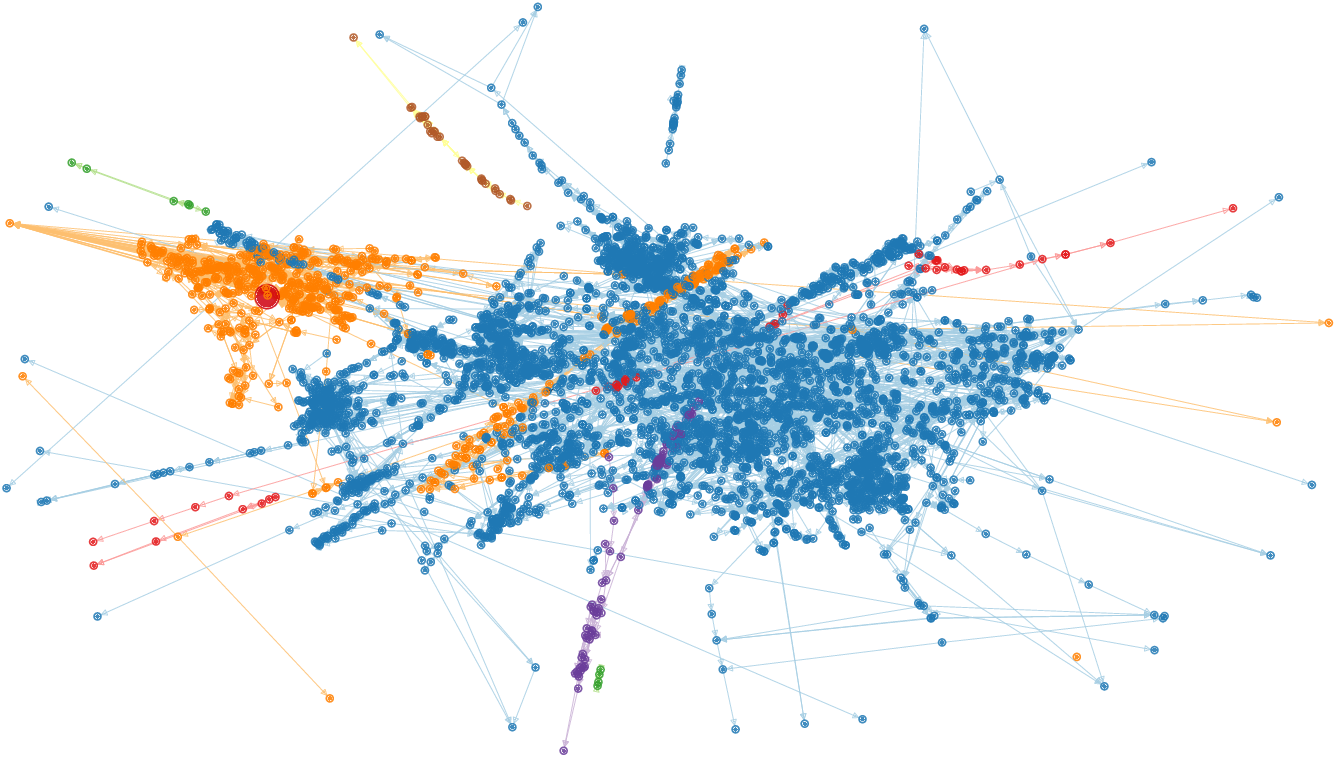}
\caption{}
\label{fig:supp-dyn3b12}
\end{subfigure}

\begin{subfigure}[c]{\textwidth}
\hspace{19ex}\includegraphics[scale=0.27]{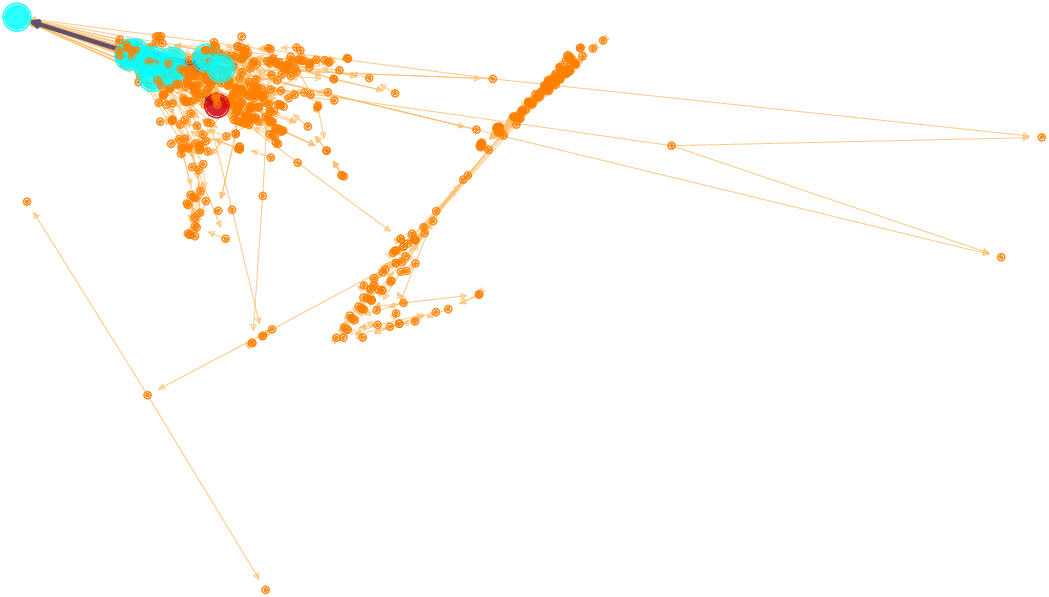}
\caption{}
\label{fig:supp-dyn3b13}
\end{subfigure}

\caption{DYN3BUG first iteration. In these figures, the large red node 
is the bug, and the large, light blue nodes designate the most central 
nodes to be sampled.}
\label{fig:supp-dyn3b-1-supp}
\end{figure*}

\begin{figure*}[th!]
\begin{subfigure}[c]{\textwidth}
\hspace{20ex}\includegraphics[scale=0.27]{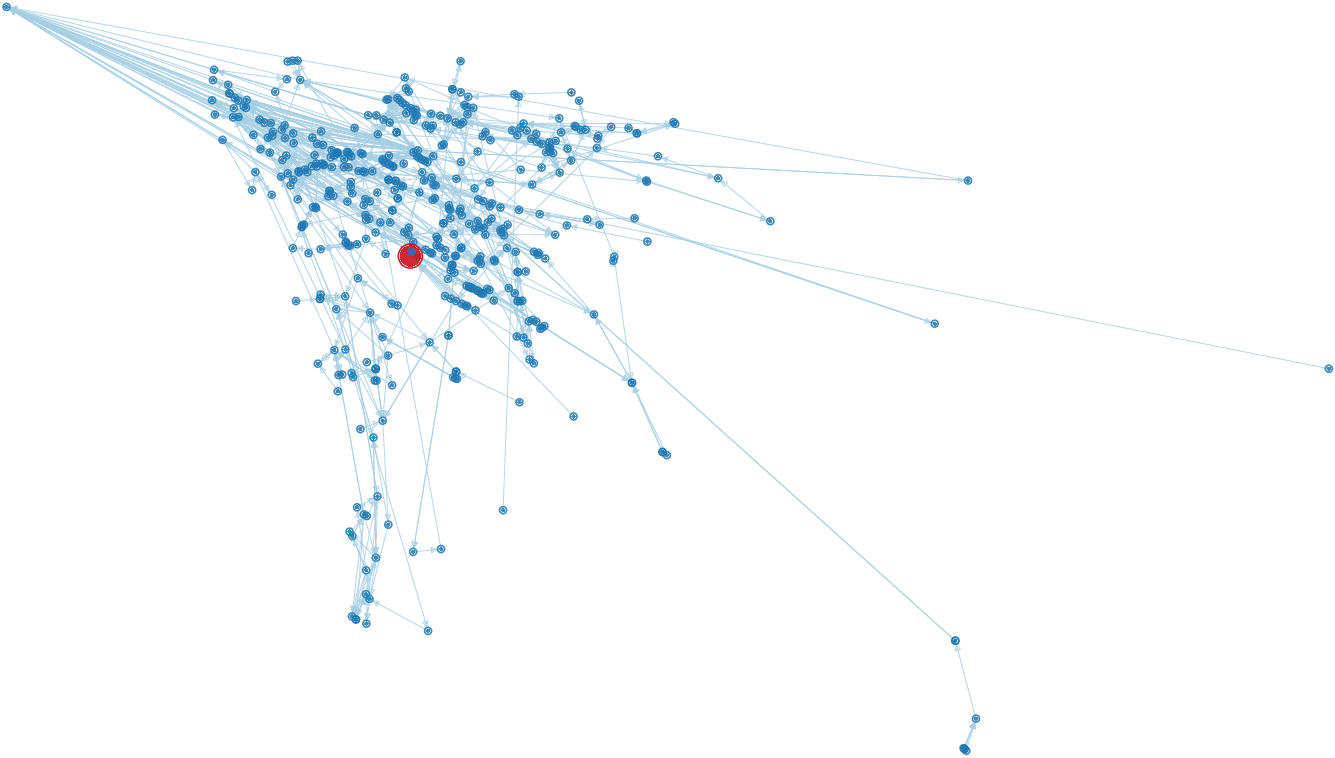}
\caption{}
\label{fig:supp-dyn3b21}
\end{subfigure}

\begin{subfigure}[c]{\textwidth}
\hspace{20ex}\includegraphics[scale=0.27]{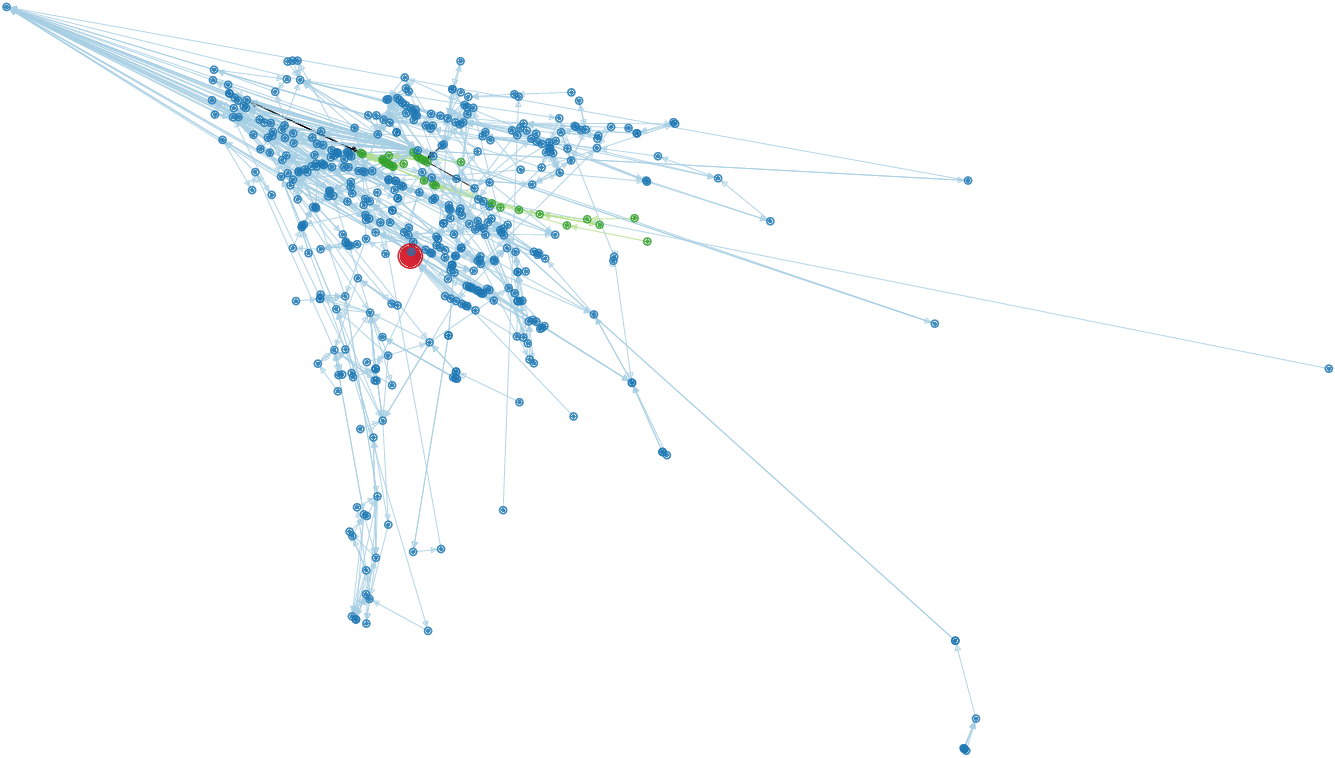}
\caption{}
\label{fig:supp-dyn3b22}
\end{subfigure}

\begin{subfigure}[c]{\textwidth}
\hspace{20ex}\includegraphics[scale=0.27]{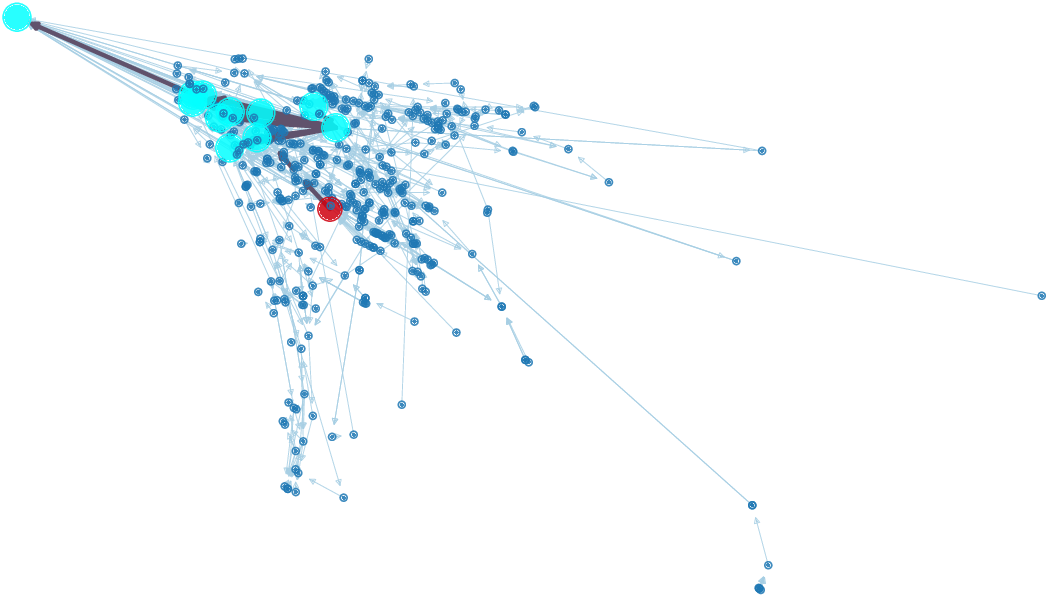}
\caption{}
\label{fig:supp-dyn3b23}
\end{subfigure}

\caption[DYN3BUG]
{DYN3BUG second iteration. In these figures, the large red node 
is the bug, and the large, light blue nodes designate the most central 
nodes to be sampled.}
\label{fig:supp-dyn3bug-supp}
\end{figure*}

Another example of a bug consisting of a single 
line change is located in a dynamics subroutine that 
computes hydrostatic pressure in the CAM core. The bug particularly affects 
the five variables listed in Table \ref{tab:supp-varselec}. 
We apply our iterative refinement to the induced subgraph of 
5,999 nodes and 11,495 edges (Figure \ref{fig:supp-dyn3b11}), and 
successfully separate the orange dynamics community from the blue 
physics community.  Instrumenting the light blue, most central nodes 
in Figure \ref{fig:supp-dyn3b13} would detect a difference in values
between ensemble and experimental runs, as at least one 
instrumented node is reachable from the bug. 
Inducing a subgraph on nodes contained in paths terminating 
on the central nodes connected to the bug further reduces 
the size of the subgraph. The second iteration of the 
refinement procedure yields a subgraph identical to 
Figure \ref{fig:supp-dyn3b21}, so further refinement will 
not be possible without analysis of true runtime values.

\subsubsection{AVX2}

\begin{figure*}[th!]
\centering
\includegraphics[width=\textwidth]{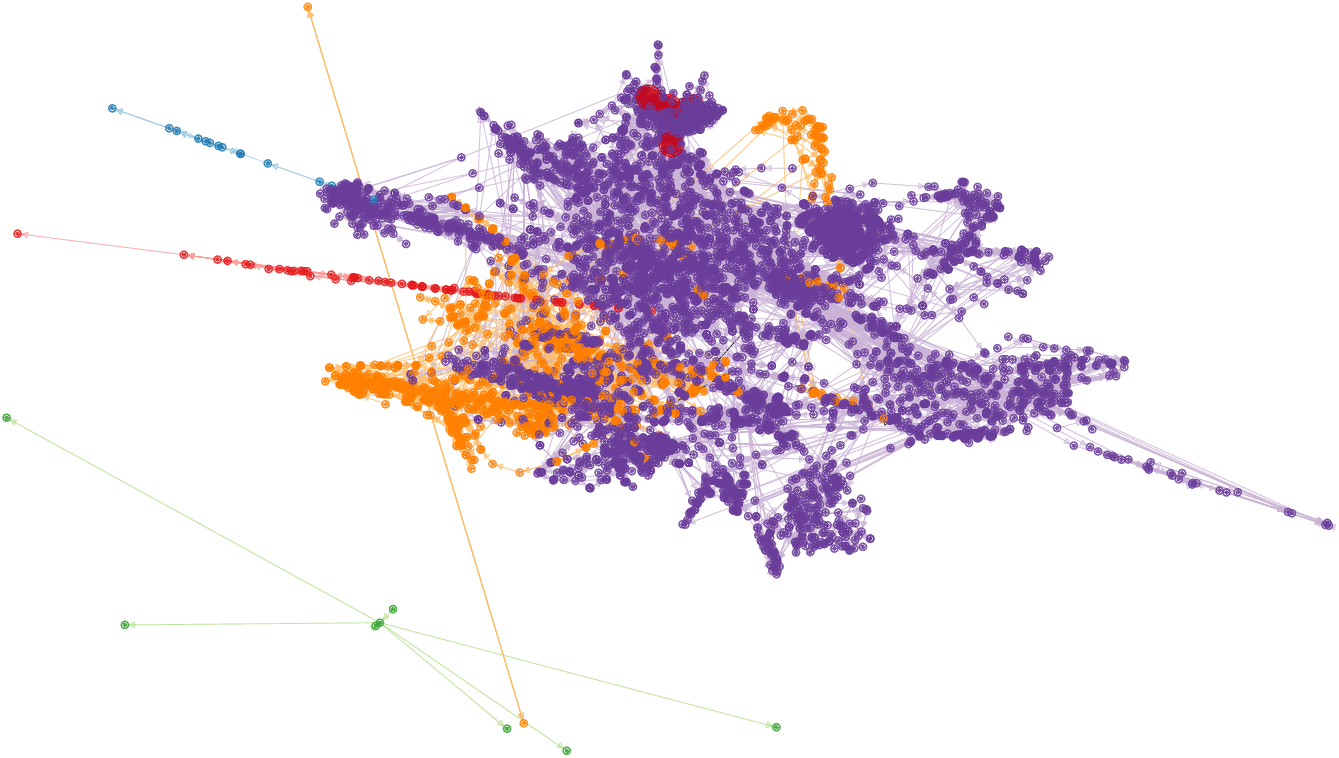}
\caption[AVX2 No models]
{Communities generated by the induced subgraph defined 
by variables affected by the AVX2 experiment. Variable locations 
are not restricted to CAM. The variables identified by KGen 
are colored red.}
\label{fig:supp-avx2-nomodels}
\end{figure*}

Figure \ref{fig:supp-avx2-nomodels} is an assertion that restricting 
our induced subgraph nodes to variables present in CAM is not necessary 
for finding the sources of inconsistency (but it does reduce the number 
of algorithm \ref{alg:graphrefinement} iterations for plotting purposes).  
This subgraph is created with the same affected variable list as Figure \ref{fig:avx2}, 
but allows nodes outside of CAM (such as in the land model). 
Although the graph is larger (7,796 nodes and 16,532 edges), it 
manifests the community structure of the CAM core (purple cluster). 
The second iteration of algorithm \ref{alg:graphrefinement} reveals a 
community that is very similar to the AVX2 subgraph in Section \ref{subsec:AVX2}. 
The nodes with the largest centralities in the community are the same 
as those in Figure \ref{fig:avx21}.
This suggests that the same conclusions are reached with this subgraph 
after a single additional of iteration.

\end{document}